\pdfoutput=1
\documentclass[preprint]{elsarticle}

\usepackage{amssymb}
\usepackage{amsmath}
\usepackage{hyperref}
\usepackage{pdflscape}
\usepackage{longtable}

\DeclareUnicodeCharacter{2032}{'}
\journal{Computational Materials Science}
\begin{document}
\begin{frontmatter}

\title{Machine learning predictions of superalloy microstructure}
\author[inst]{Patrick L. Taylor\corref{cor}}
\author[inst]{Gareth Conduit}
\affiliation[inst]{Cavendish Laboratory, University of Cambridge, CB3 0HE, UK}
\cortext[cor]{Corresponding author at: Cavendish Laboratory, University of Cambridge, CB3 0HE, UK. Email: pt409@cam.ac.uk}

\begin{abstract}
    Gaussian process regression machine learning with a physically-informed kernel is used to model the phase compositions of nickel-base superalloys. The model delivers good predictions for laboratory and commercial superalloys, with $R^2>0.8$ for all but two components of each of the $\gamma$ and $\gamma'$ phases, and $R^2=0.924$ ($\mathrm{RMSE}=0.063$) for the $\gamma'$ fraction. 
    For four benchmark SX-series alloys the methodology predicts the $\gamma'$ phase composition with $\mathrm{RMSE}=0.006$ and the fraction with $\mathrm{RMSE}=0.020$, superior to the $0.007$ and $0.021$ respectively from CALPHAD. Furthermore, unlike CALPHAD Gaussian process regression quantifies the uncertainty in predictions, and can be retrained as new data becomes available. 
\end{abstract}


\end{frontmatter}

\section{Introduction}
Nickel--base superalloys boast excellent high-temperature properties including yield stress, creep strength, fatigue life, and oxidation resistance. 
For this reason they are a critical material for the aerospace industry, as well as finding use in ground-based gas turbines, steam turbines, and nuclear reactors. 
The excellent high-temperature properties of superalloys arise due to their unique sub-crystal scale matrix/precipitate microstructure~\cite{Reed2009Alloys-By-Design:Superalloys,Reed2016Alloys-By-Design:Superalloys,Li2018InfluenceSuperalloy}. 
Accurate design of future alloys requires a detailed understanding of this underlying structure.
Consequently, first-principles modelling of superalloys is dependent on accurate prediction of both the amount and chemical composition of both the matrix ($\gamma$) and precipitate ($\gamma'$) phases. 
Two main methods are used to understand the microstructure: physical modelling and curve-fitting.

Physical modelling methods must trade off between accuracy and computational cost, with density functional theory (DFT) regarded by many as the ideal compromise~\cite{vandeWalle2002AutomatingCalculations,vandeWalle2009MulticomponentToolkit,VanDeWalle2013EfficientStructures}. Its computational cost scales as $\mathcal{O}(N^3)$ with the $N$ atoms in the simulation cell, the required size of which scales with the number of components in the alloy system~\cite{Schuch2009ComputationalFunctionaltheory}. For this reason an increasingly popular approach is to use a limited number of DFT calculations to fit the parameters of a faster atomistic model~\cite{Sanchez2010ClusterAlloys,Seko2009ClusterCalculations,Novikov2020TheLearning,Gubaev2018Machine-LearningAlloys,Rosenbrock2021Machine-learnedDiagrams,Deringer2019MachineScience}. However, the cutting-edge models are still limited to quinary alloys due to the cost of the initial DFT calculations~\cite{Nguyen2017Cluster-ExpansionMagnets,Grabowski2019AbAlloys}. 

The principle curve-fitting approach is the Calculation of Phase Diagrams (CALPHAD) methodology~\cite{Andersson2002Thermo-CalcScience,Sundman2015OpenCalphadSoftware}. CALPHAD is an equilibrium thermodynamics approach that uses a free energy model for each phase to explore equilibrium properties~\cite{Kattner2016TheDevelopment,Hillert1980EmpiricalPhases,Chang2004PhaseFuture}. This is an inverse problem approach to fitting experimental phase composition data~\cite{Bajaj2011TheMethod}. Consequently we anticipate that the advantage of CALPHAD over other curve-fitting methods is to reproduce sensible physical limits when extrapolating beyond the range of the training data.  
However, traditional CALPHAD methods do not calculate uncertainties, so users cannot understand how trustworthy a given prediction is, which limits the usefulness of their predictions. Recent work by Attari \textit{et al} attempted to address this issue by using a Markov Chain Monte Carlo approach to work out the parameter values in a CALPHAD model, and hence infer uncertainties within a Bayesian framework~\cite{Attari2020UncertaintyModel}. 
Another curve-fitting method is the Alloy Design Program developed at NIMS, Japan, by Harada \textit{et al}. 
The program calculates the microstructure of a superalloy from its nominal composition using linear regression models of the partitioning coefficients for each element, and was used to design the TMS-series superalloys~\cite{Harada1988PhaseSuperalloys,Harada1999DesignSuperalloys,Kawagishi2012DevelopmentTMS-238,Sato2008AProcessability,Enomoto1991CalculationMethod,Saito1997TheMetals}. 

Machine learning (ML) is a new approach to curve-fitting data that has recently gained prominence in materials science.
The focus has been on using ML to predict the macroscale properties of superalloys, so that the ML models can be used to design new alloy compositions to a given design criteria~\cite{Reed2009Alloys-By-Design:Superalloys,Reed2016Alloys-By-Design:Superalloys,Venkatesh1999NeuralPrediction,Yoo2002CompositionalNetwork,Tancret2003DesignModelling,Conduit2017DesignNetwork,Conduit2019ProbabilisticDeposition,Menou2016Multi-objectiveAlgorithms,Menou2019ComputationalTemperature,Liu2020PredictingLearning}. A wide range of ML approaches have been used, including neural networks and Gaussian process regression (GPR). 
Other researchers have applied ML to predicting different aspects of superalloy microstructure such as the lattice misfit~\cite{Zhang2020LatticeSuperalloys}. Yabansu \textit{et al} used GPR to predict the evolution of the microstructure morphology with ageing, taking as input the time and temperature of the ageing heat treatment~\cite{Yabansu2019ApplicationSuperalloys}. 
The ML approach is appealing as it can not only address all materials properties, but also estimate the uncertainties in its predictions~\cite{Kawagishi2012DevelopmentTMS-238,Yuan2014Creep800C}. 

In this paper we combine the best of physical-based methods and machine learning. Our approach is to develop a GPR model with a physically-informed kernel to capture the underlying physical principles. 
We first compile a training database from the freely available scientific literature. 
We then describe the framework of the GPR model used to predict the chemical composition and fractions of each phase, including a novel kernel devised to encapsulate some basic physical principles of alloy systems. The predictions of the GPR model are then initially compared to experimental values before being compared with CALPHAD results on a selection of four experimentally measured SX--series superalloys proposed in Ref.~\cite{Sulzer2020TheApplications} as an uncertainty benchmark. 

\section{Data processing}
\label{sect:database_processing}
Machine learning requires data to train on. For this reason it is necessary to compile a database that we draw from both laboratory and commercial superalloys: with each entry encapsulating information for both the features that the ML method will take as inputs and also the properties that it will predict as outputs. In this work each row in the database is a different superalloy. 
The columns are the alloy's descriptors comprising of composition $\mathbf{x}_\mathrm{c}=[x_1,...,x_n]^\mathrm{T}$ (at. \%), ageing heat treatment $\mathbf{x}_\mathrm{h}=[x_1^\mathrm{HT},...,x_{2m}^\mathrm{HT}]^\mathrm{T}$, and also its properties that are the components of both the alloy's $\gamma$ ($\mathbf{x}^\gamma=[x_i^\gamma]^\mathrm{T}$) and $\gamma'$ ($\mathbf{x}^{\gamma'}=[x_i^{\gamma'}]^\mathrm{T}$) phases and also the $\gamma'$ phase fraction $f$. 
Since the composition must sum to one, a reduced description $\mathbf{x}_\mathrm{c} \mapsto \bar{\mathbf{x}}_\mathrm{c}=[x_2,...,x_n]^\mathrm{T}$ could be used. However principal component analysis (Fig.~\ref{fig:PCA_comparison}) showed that low rank projections of $\mathbf{x}_\mathrm{c}$ had more explanatory power than the same rank projections of $\bar{\mathbf{x}}_\mathrm{c}$ (blue and red line respectively in Fig.~\ref{fig:PCA_comparison}). For this reason the full set of descriptors were used throughout. 
We focus on pure $\gamma/\gamma'$ alloys and therefore circumvent consideration of any carbides and secondary phases that could form~\cite{Reed2016Alloys-By-Design:Superalloys,Reed2006TheApplications,Durand-Charre1997TheSuperalloys,Wang2008RoleSuperalloys,Zhang2019DesignValidations,Probstle2016ImprovedElements,Hobbs2008TheSuperalloys}. 

Heat treatments are applied to superalloys to control the microstructure morphology, and can be classed as either a solution heat treatment or a precipitation/ageing heat treatment. 
The solution heat treatment homogenises the microstructure so its temperature and duration is dictated by the superalloy composition~\cite{Durand-Charre1997TheSuperalloys,Caron2008InfluenceSuperalloys,Khan1984TheSuperalloy}. 
Therefore, we consider the solution heat treatment as a property of the overall alloy composition, rather than as an input parameter. 
Next, ageing heat treatments are applied to further optimise the morphology of the superalloy microstructure. 
Six features describe the ageing heat treatment: a heat treatment temperature and time, for up to three heat treatment stages per alloy. For alloys with fewer total stages, the remaining ones are specified as being at room temperature for no time.

The database comprises 97 experimental superalloy microstructures, of which 36 entries correspond to laboratory alloys, 43 to commercial (or modifications of) single-crystal superalloys, and 18 to other commercial polycrystalline superalloys.
All of these entries contained both Ni and Al. Other elements occurred in a subset of database entries, between 86-times for Cr to just 5-times for Ru~\cite{Harada1988PhaseSuperalloys,Sulzer2020TheApplications,Durand-Charre1997TheSuperalloys,Khan1984TheSuperalloy,Duval1994PhaseMC2,Glas1996OrderSuperalloys,Harada1993Atom-probeSuperalloy,Khan1986EffectCMSX-2,Miller1994APFIMSuperalloy,Royer1998InSuperalloy,Diologent2004OnSuperalloys,Segersall2015Thermal-mechanicalAlloying,Schmidt1992Effect99,Nathal1985Elevated100,Loomis1972TheSuperalloys,Morrow1975TheAlloys,Pettinari2002StackingSuperalloys,Ofori2004ASuperalloys,Tin2004AtomicSuperalloys,Delargy1983PhaseIN939,Ralph1982TheTechniques,Blavette1996Atomic-scaleSuperalloys,Fuchs2002ModelingSuperalloy,Yoon2007EffectsObservations,Parsa2015AdvancedSuperalloys,Sieborger2001TemperaturePhases,Fahrmann1999DeterminationRafting,Yuan2011InfluenceSuperalloys,Ma2007DevelopmentSuperalloys,Cui2012DynamicEnergy,Jovanovic1998MicrostructureSuperalloy,Wang2014TheTemperatures,Collier1986EffectsProperties.,Long2018MicrostructuralReview,Wlodek1996TheDT,Lapington2018CharacterizationSuperalloys}. See Table \ref{tab:data_summary} for a full overview including summary statistics. 
In order to make the training procedure as robust and user-friendly as possible, no outliers were filtered from the dataset before use in the ML model. 
\begin{landscape}
\centering

\begin{table}
\begin{tabular}{lrrrrrrrrrrrr}
  & \multicolumn{12}{c}{Descriptors} \\ 
  \cline{2-13}\noalign{\vskip\doublerulesep\vskip-\arrayrulewidth}\cline{2-13}
 & \multicolumn{12}{c}{Phase composition (at. \%)} \\ 
 & Ni & \multicolumn{1}{c}{Cr} & \multicolumn{1}{c}{Co} & \multicolumn{1}{c}{Re} & \multicolumn{1}{c}{Ru} & \multicolumn{1}{c}{Al} & \multicolumn{1}{c}{Ta} & \multicolumn{1}{c}{W} & \multicolumn{1}{c}{Ti} & \multicolumn{1}{c}{Mo} & \multicolumn{1}{c}{Nb} & \multicolumn{1}{c}{Hf} \\ \hline
Min ($>0$) & 47.28 & 3.00 & 2.50 & 0.64 & 1.30 & 2.00 & 0.44 & 0.03 & 1.12 & 0.31 & 0.006 & 0.03 \\
Max & 86.50 & 24.47 & 18.80 & 2.50 & 3.50 & 14.20 & 4.03 & 5.76 & 5.84 & 5.16 & 1.20 & 0.33 \\
Median & 66.50 & 12.55 & 6.65 & 2.00 & 3.50 & 11.20 & 1.99 & 1.53 & 2.70 & 1.30 & 0.44 & 0.06 \\
Mean & 67.46 & 12.30 & 8.20 & 1.75 & 3.06 & 9.89 & 1.97 & 1.83 & 2.88 & 1.97 & 0.39 & 0.13 \\
Frequency & 97 & \multicolumn{1}{c}{96} & \multicolumn{1}{c}{59} & \multicolumn{1}{c}{18} & \multicolumn{1}{c}{5} & \multicolumn{1}{c}{97} & \multicolumn{1}{c}{55} & \multicolumn{1}{c}{54} & \multicolumn{1}{c}{52} & \multicolumn{1}{c}{61} & \multicolumn{1}{c}{15} & \multicolumn{1}{c}{7} \\ \hline
 & \multicolumn{6}{c}{Heat treatment ($^\circ$C $\lvert$ hrs)} &  &  &  &  &  &  \\ 
 & \multicolumn{2}{c}{\#1} & \multicolumn{2}{c}{\#2} & \multicolumn{2}{c}{\#3} & \multicolumn{2}{r}{} &  &  &  &  \\ \cline{1-7}
Min ($>0$) & \multicolumn{1}{r}{760} & 0.25 & 760 & 0.25 & 700 & 0.5 &  &  &  &  &  &  \\
Max & \multicolumn{1}{r}{1300} & 1500 & 1160 & 264 & 1050 & 100 &  &  &  &  &  &  \\
Median & \multicolumn{1}{r}{980} & 8 & 850 & 24 & 1040 & 16 &  &  &  &  &  &  \\
Mean & 963.8 & 305.2 & 856.9 & 31.0 & 941.4 & 24.8 & \multicolumn{1}{l}{} & \multicolumn{1}{l}{} & \multicolumn{1}{l}{} & \multicolumn{1}{l}{} & \multicolumn{1}{l}{} & \multicolumn{1}{l}{} \\
Frequency & \multicolumn{2}{c}{92} & \multicolumn{2}{c}{38} & \multicolumn{2}{c}{7} &  &  &  &  &  &  \\ \cline{1-7}
 \vspace{2pt} \\
 & \multicolumn{12}{c}{Properties} \\ 
 \cline{2-13}\noalign{\vskip\doublerulesep\vskip-\arrayrulewidth}\cline{2-13}
 & \multicolumn{11}{c}{log partitioning coefficient $P_i$} & \multicolumn{1}{c}{} \\ 
 & Ni & \multicolumn{1}{c}{Cr} & \multicolumn{1}{c}{Co} & \multicolumn{1}{c}{Re} & \multicolumn{1}{c}{Ru} & \multicolumn{1}{c}{Al} & \multicolumn{1}{c}{Ta} & \multicolumn{1}{c}{W} & \multicolumn{1}{c}{Ti} & \multicolumn{1}{c}{Mo} & \multicolumn{1}{c}{Nb} & \multicolumn{1}{c}{} \\ \cline{1-12}
Min & \multicolumn{1}{r}{-0.692} & 0.404 & -0.208 & 0.142 & 1.169 & -6.984 & -5.711 & -0.411 & -5.220 & -0.463 & -1.872 &  \\
Max & \multicolumn{1}{r}{0.133} & 3.086 & 1.665 & 7.078 & 1.508 & -0.488 & 0.489 & 1.329 & -0.340 & 2.006 & 1.454 &  \\
Median & \multicolumn{1}{r}{-0.154} & 2.038 & 1.077 & 1.794 & 1.218 & -1.500 & -1.892 & 0.311 & -1.740 & 1.239 & 0.022 &  \\
Mean & -0.170 & 1.926 & 1.027 & 2.640 & 1.267 & -1.669 & -2.337 & 0.355 & -2.049 & 1.021 & -0.198 & \multicolumn{1}{l}{} \\
Frequency & \multicolumn{1}{r}{97} & 96 & 59 & 18 & 5 & 97 & 54 & 52 & 52 & 59 & 13 &  \\ \hline
 & \multicolumn{11}{c}{log partitioning coefficient $P_i'$} & \multicolumn{1}{c}{$\gamma'$} \\ 
 & Ni & \multicolumn{1}{c}{Cr} & \multicolumn{1}{c}{Co} & \multicolumn{1}{c}{Re} & \multicolumn{1}{c}{Ru} & \multicolumn{1}{c}{Al} & \multicolumn{1}{c}{Ta} & \multicolumn{1}{c}{W} & \multicolumn{1}{c}{Ti} & \multicolumn{1}{c}{Mo} & \multicolumn{1}{c}{Nb} & \multicolumn{1}{c}{frac.} \\ \hline
Min & \multicolumn{1}{r}{-0.297} & 0.274 & -0.037 & 0.122 & 0.390 & -1.410 & -1.208 & -0.446 & -1.293 & -0.405 & -0.795 & 0.030 \\
Max & \multicolumn{1}{r}{0.128} & 2.610 & 1.145 & 3.039 & 0.722 & 0.621 & 0.666 & 1.116 & -0.039 & 1.813 & 0.774 & 0.776 \\
Median & \multicolumn{1}{r}{-0.058} & 1.320 & 0.508 & 0.857 & 0.568 & -0.445 & -0.356 & 0.100 & -0.423 & 0.594 & -0.432 & 0.470 \\
Mean & -0.063 & 1.361 & 0.550 & 1.194 & 0.565 & -0.518 & -0.371 & 0.173 & -0.550 & 0.643 & -0.136 & 0.450 \\
Frequency & 97 & \multicolumn{1}{c}{96} & \multicolumn{1}{c}{59} & \multicolumn{1}{c}{15} & \multicolumn{1}{c}{5} & \multicolumn{1}{c}{97} & \multicolumn{1}{c}{54} & \multicolumn{1}{c}{52} & \multicolumn{1}{c}{52} & \multicolumn{1}{c}{59} & \multicolumn{1}{c}{8} & \multicolumn{1}{c}{97} \\ \hline
\end{tabular}
\caption{Summary of the database.}
\label{tab:data_summary}
\end{table}

\end{landscape}

\subsection{Data parameterisation}
With the data curated we are now well-positioned to reparameterise it to both capture underlying physics and also to make it more amenable to machine learning. 
To express the phase composition we adopt logarithmic partitioning coefficients $P_i$, $P_i'$~\cite{Harada1999DesignSuperalloys,Harada1988PhaseSuperalloys,Ofori2004ASuperalloys,Tin2004AtomicSuperalloys,Fuchs2002ModelingSuperalloy,Ma2007DevelopmentSuperalloys,Wang2014TheTemperatures}. They are defined as:
\begin{align}
    P_i = \log\left(\frac{x^\gamma_i}{x^{\gamma'}_i}\right)\ , \qquad P'_i=\log\left(\frac{x_i}{x^{\gamma'}_i}\right) \quad. \label{eq:part_coeff_def}
\end{align}
Predicting the logarithm of the partitioning coefficients ensures they always have physical positive values and preserves the symmetry between the $\gamma$ and $\gamma'$ phases.

A superalloy comprises $n$ elements and $p$ phases, giving a total of $np+p$ variables of interest. The goal is to calculate them from the nominal composition of the superalloy $\mathbf{x}=[x_1,...,x_n,x_1^\mathrm{HT},...,x_{2m}^\mathrm{HT}]^\mathrm{T}$. There are also the following physical constraints on these properties:
\begin{align}
    &\begin{aligned}&\textrm{Total components sum to unity for}\\
    &\textrm{each phase.}\end{aligned} 
    && \sum_i x^\alpha_i = 1 \label{eq:comp_constraint} \\
    &\textrm{Total phases sum to unity.} && \sum_\alpha f_\alpha = 1 \label{eq:frac_constraint} \\
    &\begin{aligned}&\textrm{Sum of all elements in all phases is}\\
    &\textrm{amount of element in the material.}\end{aligned} 
    && \sum_\alpha f_\alpha x^\alpha_i = x_i \label{eq:comp_physical_constraint}
\end{align}
In our database entries $f$ is often determined from the measured chemical composition of each phase using the $n$ versions of Eq.~\ref{eq:comp_physical_constraint}, and is hence algebraically over-determined, resulting in an approximate value determined by finding the best fit for $f$ to a rearrangement of Eq.~\ref{eq:comp_physical_constraint}  \cite{Segersall2015Thermal-mechanicalAlloying,Reed2004IdentificationTomography}. Typically this means Eq.~\ref{eq:comp_physical_constraint} does not hold exactly.  Another reason this equation may not be fulfilled exactly is that the $\gamma'$ phase fraction has been measured not only independently of the phase compositions but also via another method---e.g. via atom probe tomography or chemical analysis for the phase composition versus an ocular determination from SEM imagery for the phase fraction. In both cases Eq.~\ref{eq:comp_physical_constraint} should hold to within the experimental tolerances of the independent $\gamma$ and $\gamma'$ phase composition and fraction measurements.

Relaxing Eq.~\ref{eq:comp_physical_constraint} from a strict constraint means that we need to predict all but one phase fraction, and for each phase all but one component, i.e. a total of $np-1$ predictions. In the subsequent sections we will refer to the specific case of superalloys, with $p=2$ phases. $f$ will refer to the fraction of $\gamma'$ or precipitate phase in the alloy, and $f^\gamma$ can then be inferred by by Eq.~\ref{eq:frac_constraint}. The constraints implied by Eq.~\ref{eq:comp_physical_constraint} are revisited in section \ref{subsect:output}.

\section{Computational method}
\label{sect:ml_model_for_ms}
\subsection{Machine learning methodology}
\label{subsect:ml}
We adopt Gaussian process regression (GPR) machine learning to predict the log partitioning coefficients $\hat{P}_i$ and $\hat{P}'_i$ and the $\gamma'$ phase fraction $\hat{f}$ (and corresponding uncertainties) of each element $i$ in the alloy from its features. We can then calculate the final values for the microstructure to minimise the overall uncertainty. 
GPR takes a Bayesian approach to ML, in which an optimal posterior distribution---from which predictions are made---is formed from a prior distribution and a likelihood calculated from the training dataset. The prior distribution is assumed to be Gaussian with covariance determined by a kernel function. The functional form of the kernel should be chosen to suit the problem, but its hyperparameters are optimised using the training data. In this work we follow the standard approach of maximising the log marginal likelihood of the training data~\cite{Rasmussen2006GaussianLearning,Kanagawa2018GaussianEquivalences}. 
The GPR implementation in the Scikit-learn library for Python was used~\cite{Pedregosa2011Scikit-learn:Python}. 

Specifying the kernel provides an opportunity to incorporate our prior physical knowledge into the machine learning, which should give a more accurate model with less data. Superalloy properties are usually determined mainly by their composition rather than their heat treatment~\cite{Khan1986EffectCMSX-2,Diologent2004OnSuperalloys,Schmidt1992Effect99,Ralph1982TheTechniques,Yoon2007EffectsObservations}. We capture this physical rule of thumb in the following kernel:
\begin{align}
    k(\mathbf{x},\mathbf{x}')= k_{\mathrm{comp},0}(\mathbf{x}_\mathrm{c},{\mathbf{x}_\mathrm{c}}') 
    +k_{\mathrm{comp},1}(\mathbf{x}_\mathrm{c},{\mathbf{x}_\mathrm{c}}')\ k_\mathrm{HT}(\mathbf{x}_\mathrm{h},{\mathbf{x}_\mathrm{h}}') \label{eq:kernel_scheme_2} \quad .
\end{align}
The first kernel term captures the bulk of the variation due to composition. The second kernel term captures the heat treatment and 
can be interpreted as an AND operation for the two measures of similarity, coupling heat treatment and composition~\cite{Duvenaud2014AutomaticProcesses}. 
A variation of Eq.~\ref{eq:kernel_scheme_2} in which the second composition term was a constant was also tested. 

\begin{figure}[h]
    \centering
    \includegraphics[width=4in]{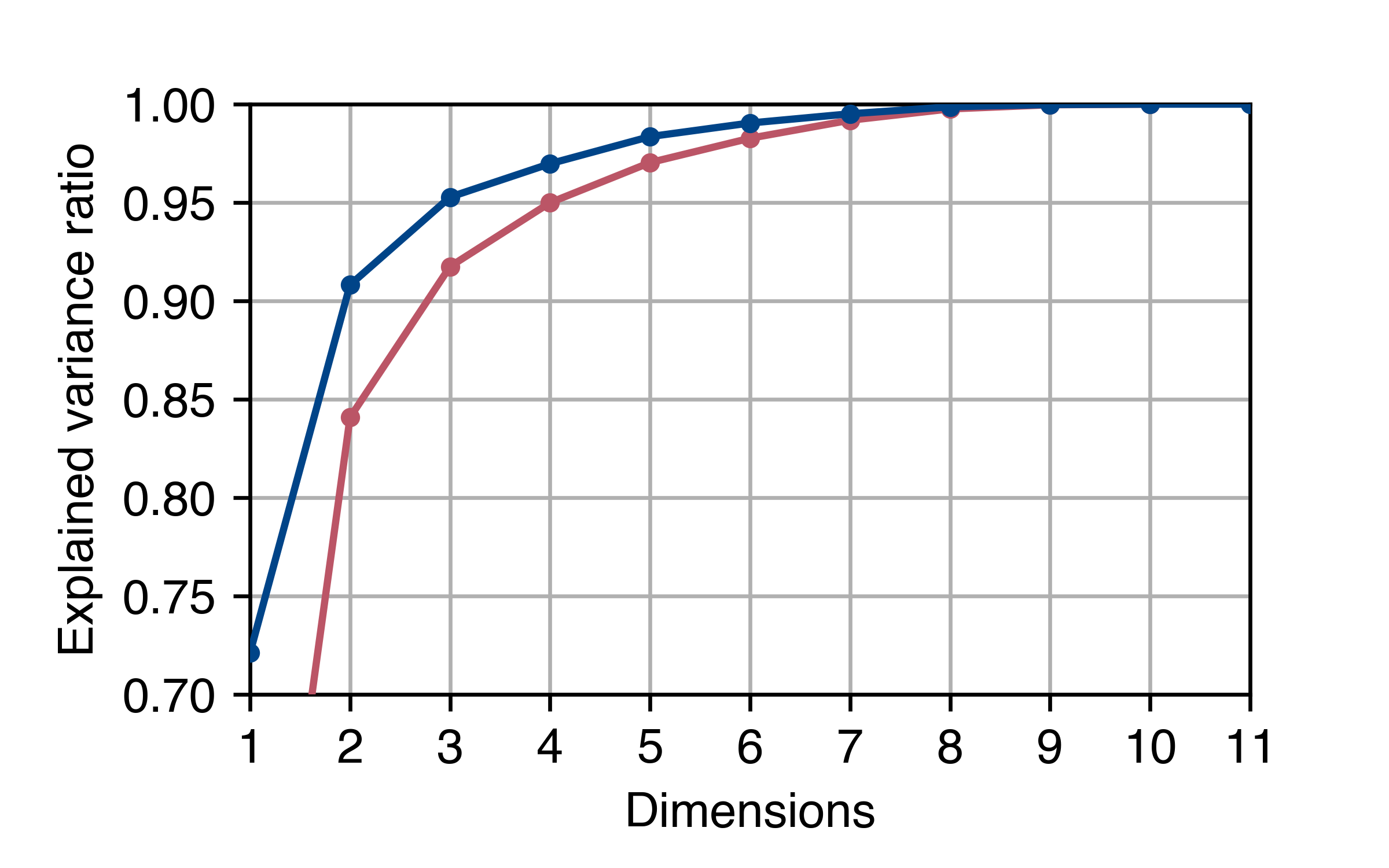}
    \caption{Explained variance with the number of dimensions by principal component analysis for the composition of superalloys included in this work. The blue line is for the full composition and the red line is for the composition excluding the base element.}
    \label{fig:PCA_comparison}
\end{figure}

Three alternate kernel functions were considered in this work. A simple and popular choice of kernel function is the Gaussian radial basis function (RBF):
\begin{align}
    k(\mathbf{x},\mathbf{x}')&=\mathrm{e}^{-{\|\mathbf{x}-\mathbf{x}'\|}^2/l^2} \label{eq:gaussian_rbf_kernel}
    \quad .
\end{align}
Here $\|\cdot\|$ is the $\ell_2$ norm. The kernel has a single hyperparameter to be optimised during fitting: the lengthscale $l$. 
Our second choice is the linear kernel:
\begin{align}
    k(\mathbf{x},\mathbf{x}')=b+\mathbf{x}\cdot\mathbf{x}' \quad , \label{eq:linear_kernel}
\end{align}
collapsing the method to ridge regression when $b=0$~\cite{Rasmussen2006GaussianLearning,Duvenaud2014AutomaticProcesses}. This was used as a comparison to the models with more complex kernels. 
The third choice is motivated by previous work that used GPR to model the effects of alloy heat treatments, which used the automatic relevance determination (ARD) variation on the Gaussian RBF kernel~\cite{Tamura2021MachineSuperalloy,Yabansu2019ApplicationSuperalloys}. This kernel introduces a length parameter for each feature, $k(\mathbf{x},\mathbf{x}')=\exp(-{\|(\mathbf{x}-\mathbf{x}')\circ \mathbf{l}^{-1}\|}^2)$.

Often the phase behaviour of an alloy is governed by the concentrations of a few elements. This low dimensional structure can be encoded in the kernel by projecting the feature vector to a lower dimensional subspace, 
$\mathbf{x}\mapsto B\mathbf{x}$, 
with $B$ being a low rank matrix. $B$ could be found prior to fitting the kernel via principal component analysis, see Fig.~\ref{fig:PCA_comparison}, or it could be optimised as a hyperparameter of the kernel during training~\cite{Snelson2012VariableProcesses}. Prior knowledge tells us that we anticipate two main groups of elements in a superalloy---Ni-like elements, and Al-like elements---with a possible third group being the refractory elements~\cite{Reed2006TheApplications,Wang2008RoleSuperalloys,Ofori2004ASuperalloys,Tin2004AtomicSuperalloys,Pollock2006Nickel-basedProperties}, suggesting that a matrix of around rank 3 would be suitable.

\subsubsection{Evaluating the quality of uncertainty prediction}
\label{subsubsect:DUQ}
GPR delivers uncertainty $\sigma_\mathrm{y}$ estimates for each of its predictions $\hat{y}$, so evaluating how close the uncertainties are to the real-life validation data, $y$, is crucial. 
The typical error in a prediction $\hat{y}-y$ should follow a Gaussian distribution with standard deviation $\sigma_y$. It follows that $(\hat{y}-y)/\sigma_y\sim\mathcal{N}(0,1)$. This quantity is accumulated over all predictions and then is arranged into a histogram and compared to an ideal ``Gaussian" histogram with $M$ bins of equal area $A=1/M$ such that each bin has equal noise. From this a distribution of uncertainty quality (DUQ) is defined as the absolute sum of the difference in the areas of bins between the two histograms. For $N$ data points this is defined as:
\begin{align}
    \mathrm{DUQ} = \frac{1}{2}\frac{M}{M-1}\sum_{m=1}^M\left\lvert \frac{n_m}{N}-\frac{1}{M}\right\rvert \quad , \label{eq:uncertainty_quality}
\end{align}
where the number of data points in each real bin, $n_m$, is normalised such that $\sum n_m = N$. The number of bins is then optimised to achieve minimal value as described in the appendix. A minimal $\mathrm{DUQ}=0$ indicates a perfect estimate of the uncertainties whereas a peak $\mathrm{DUQ}=1$ indicates a poor estimate of uncertainty.

\subsection{Calculation of phase compositions and fractions}
\label{subsect:output}
\subsubsection{Dynamic choice of balance element}
\label{subsubsect:bal_element}
For each $n$ component superalloy the relevant GPR models will predict all $2n$ partitioning coefficients $\{\hat{P}_1,...,\hat{P}_n,\hat{P}'_1,...,\hat{P}'_n\}$ and the $\gamma'$ fraction, yielding compositions $\hat{x}^\alpha_i$ and also $\hat{f}$. This is more than the total number of properties needed, meaning for each phase the amount of one element, $\hat{x}^\alpha_i$, will be calculated using Eq.~\ref{eq:comp_constraint}---which we will call the balance element, ${\hat{x}}^\alpha_{\mathrm{bal},i}$. Rather than fixing one component to be the balance element in every alloy, we instead dynamically choose whichever minimises the difference between the component's balance value and the GPR value, accounting for the uncertainty prediction of the GPR model $\sigma^\alpha_i$, that is:
\begin{align}
    \min_i\left[\frac{\left\lvert{\hat{x}}^\alpha_{\mathrm{bal},i}-\hat{x}^\alpha_i\right\rvert}{\sigma^\alpha_i}\right]
    \quad . \label{eq:comp_adjustment_choice}
\end{align}
Since the numerator is the same for every element in a given alloy, this is equivalent to choosing the element with maximal uncertainty in its value to be the balance element. 

For the final model presented in the next section, the dynamic choice of balance element was compared to a static choice, in this case the conventional choice of Ni. For Ni itself---a crucial element---the static method gave $R^2=0.757$ and $0.888$ in the $\gamma'$ and $\gamma$ phases compared with an improvement of $R^2=0.824$ and $0.927$ respectively for the dynamic method. For Al there was also an improvement with the static method giving $R^2=0.631$ ($\gamma'$) and $0.510$ ($\gamma$) compared with $R^2=0.674$ and $0.599$ for the dynamic method. This was despite Al being chosen as the balance element for 47 and 15 predictions (out of 97) for the $\gamma'$ and $\gamma$ phase respectively using the dynamic method. 

\subsubsection{Bayesian inferral of a consistent precipitate fraction}
\label{subsubsect:final_f_value}
\begin{figure}[p]
    \centering
    \includegraphics[width=4.8in]{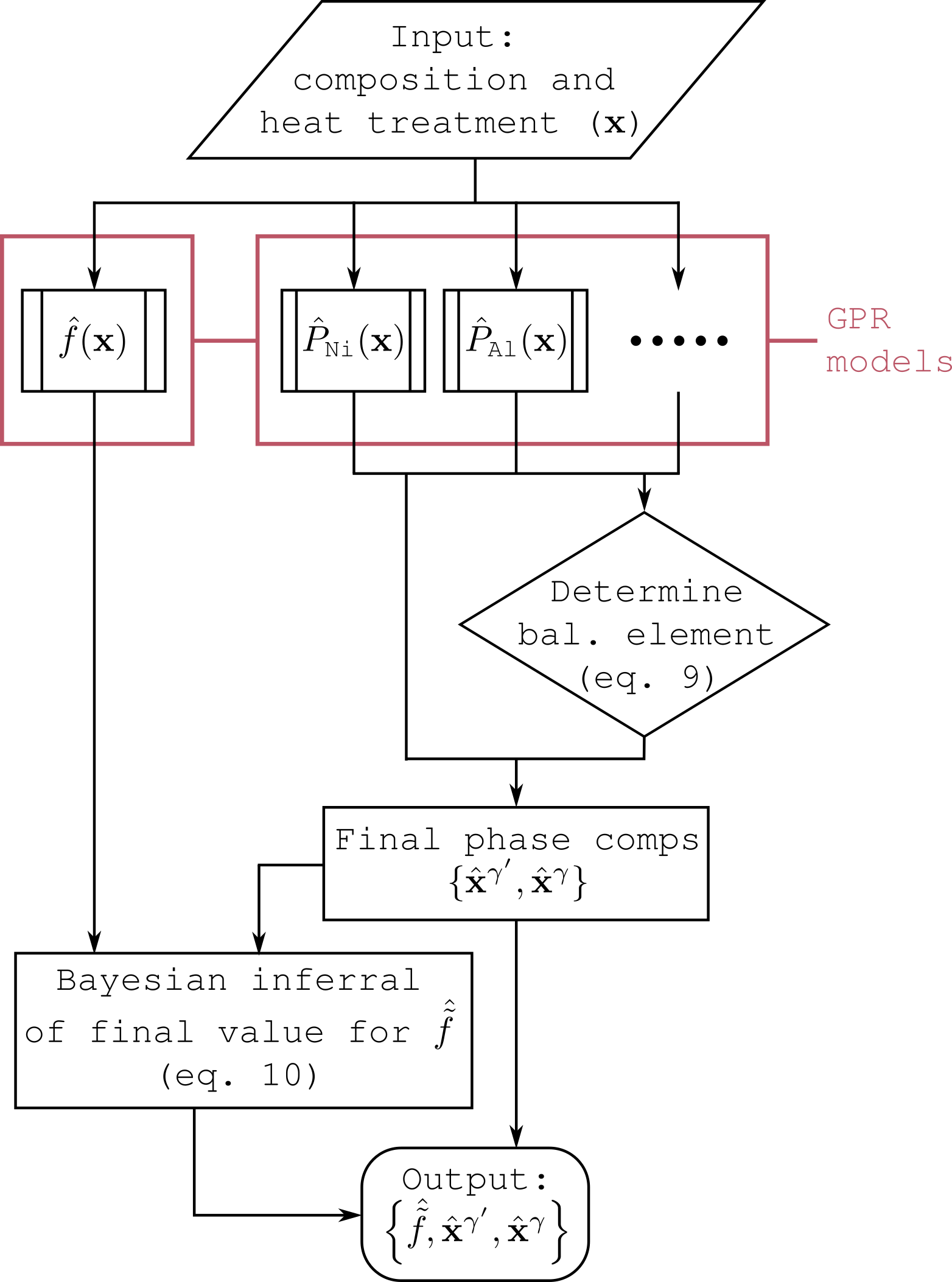}
    \caption{Flowchart outlining how microstructure predictions are made for a given alloy composition.}
    \label{fig:flowchart}
\end{figure}

Predicting the precipitate fraction $\hat{f}$, and the phase compositions $\hat{\mathbf{x}}^\gamma$ and $\hat{\mathbf{x}}^{\gamma'}$ gives us a full determination of the microstructure. 
However the composition  for each phase may not sum to the total composition---that is Eq.~\ref{eq:comp_physical_constraint} will not hold exactly but it should hold approximately~\cite{Segersall2015Thermal-mechanicalAlloying,Reed2004IdentificationTomography}. To improve consistency a Bayesian approach was taken to synthesise the information from the output of the ML models for the phase compositions and for $f$ to give the phase composition most likely to be consistent with a valid total composition. 
The output of the GPR model for $f$ is taken as a prior, and the likelihood is taken to be:
\begin{align*}
    P\left(\mathbf{x}|\hat{\mathbf{x}}^\gamma,\hat{\mathbf{x}}^{\gamma'},f,\sigma_i\right)
    \propto 
    \prod_i \exp\left[-{\frac{\left(x_i-f\hat{x}_i^{\gamma'}-(1-f)\hat{x}_i^\gamma\right)^2}{2\sigma_i^2}}\right]
    \quad .
\end{align*}
With the standard deviation $\sigma_i$ calculated from the uncertainties on the compositions as $\sigma_i^2={\hat{f}}^2{\sigma_i^{\gamma'}}^2+(1-\hat{f})^2{\sigma_i^{\gamma}}^2$ we have a conjugate prior to the likelihood and hence the posterior is also a Gaussian, with mean value and standard deviation:
\begin{align}
    \hat{\tilde{f}} &= \frac{\hat{f}+
    \sum_i\Big(x_i-\hat{x}_i^\gamma\Big)\Big(\hat{x}_i^{\gamma'}-\hat{x}_i^\gamma\Big)
    \left(\frac{\sigma_{\mathrm{f}}}{\sigma_i}\right)^2}
    {1+\sum_i\left(\hat{x}_i^{\gamma'}-\hat{x}_i^\gamma\right)^2
    \left(\frac{\sigma_{\mathrm{f}}}{\sigma_i}\right)^2} \label{eq:calc_f} \\
    \tilde{\sigma}_\mathrm{f}^2 &= \frac{\sigma_{\mathrm{f}}^2}{1+\sum_i\left(\hat{x}_i^{\gamma'}-\hat{x}_i^\gamma\right)^2
    \left(\frac{\sigma_{\mathrm{f}}}{\sigma_i}\right)^2} \label{eq:calc_f_std} \quad ,
\end{align}
which are taken as the final values for the precipitate fraction and its uncertainty. Note that in the limit $\sigma_{f}\gg\sigma_i $, Eq.~\ref{eq:calc_f} agrees with the method of Reed et al~\cite{Segersall2015Thermal-mechanicalAlloying,Reed2004IdentificationTomography}. 
The method described above can be viewed as transforming Eq.~\ref{eq:comp_physical_constraint} from a strict constraint into a ``soft" probabilistic constraint. 
A flowchart overview of how the final phase compositions and fractions are predicted for a given input composition is shown in Fig.~\ref{fig:flowchart}. 

\section{Results and discussion}
\label{sect:ml_ms_model_results}
\begin{figure}[h]
    \centering
    \includegraphics[width=4in,
    clip]{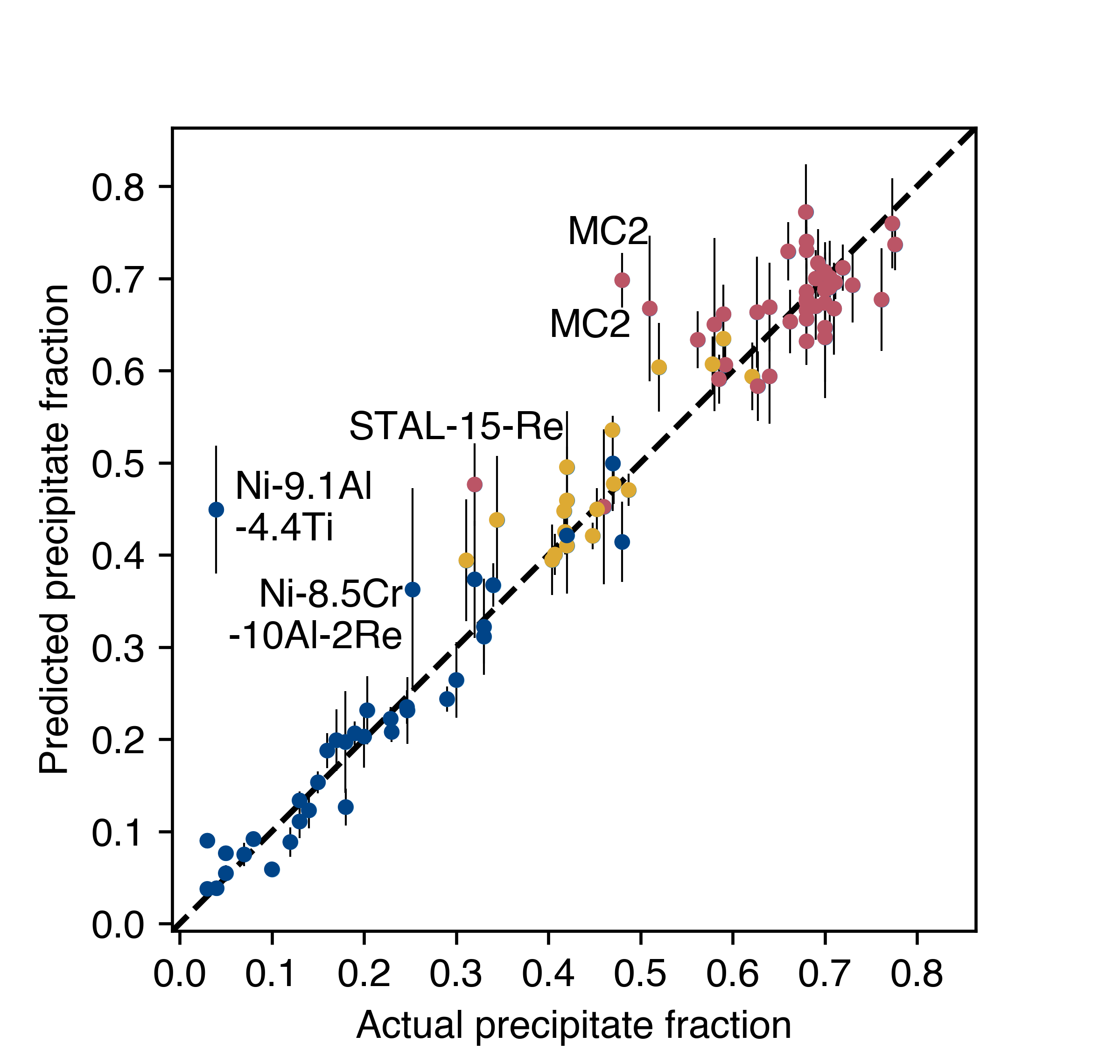}
    \caption{Predicted precipitate ($\gamma'$ phase) fraction with error bars for validation data from GPR models fitted across 5-folds, with kernel scheme Eq.~\ref{eq:kernel_scheme_2} and functional form Eq.~\ref{eq:good_kernel_2}. Colours refer to different types of superalloy: blue are laboratory alloys, yellow are other polycrystalline superalloys, and red are single-crystal superalloys. We highlight the composition of one particular superalloy. 
    }
    \label{fig:frac_final}
\end{figure}
\begin{figure}[h]
    \centering
    \includegraphics[width=4in,
    clip]{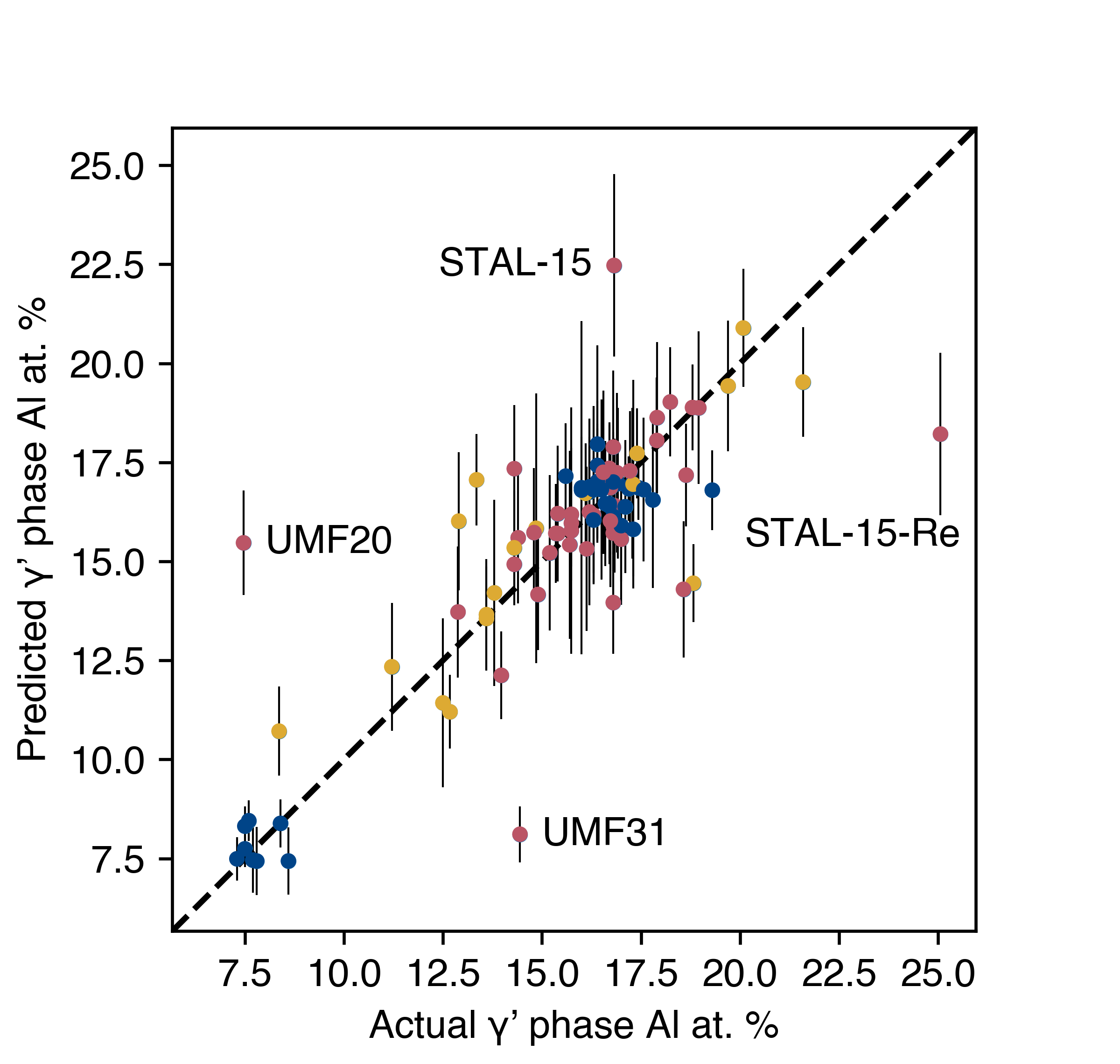}
    \caption{Predicted Al at. \% in the $\gamma'$ phase for validation data from GPR models fitted across 5-folds, with kernel scheme Eq.~\ref{eq:kernel_scheme_2} and functional form Eq.~\ref{eq:good_kernel_2}. Colours refer to different types of superalloy: blue are lab alloys, yellow are other polycrystalline superalloys, and red are single-crystal superalloys. 
    }
    \label{fig:Al_in_detail}
\end{figure}

Variations of the GPR model with different kernels were trained on the database described in section \ref{sect:database_processing}. 
For each GPR model 5-fold cross validation was carried out. This procedure was used to fairly assess the quality of the different models in a way that will also highlight possible issues with over-fitting owing to a large number of variables. The step-by-step procedure is:
\begin{enumerate}
    \item 80\% of the initial dataset it selected at random as \textit{training} data, with the remaining 20\% being \textit{validation} data. 
    \item A GPR model is trained with the training data without sight of the validation data. Here the training procedure is to fit the kernel hyperparameters so as to maximise the log marginal likelihood of the training data~\cite{Rasmussen2006GaussianLearning,Kanagawa2018GaussianEquivalences}.
    \item The GPR model is tested against the validation data to get a measure of the accuracy. If the training data was over-fitted owing to a large number of model variables then the model will deliver poor accuracy against the blind validation data. Similarly, if the model has not captured the underlying physics of the dataset, it will perform comparatively poorly on the validation data.
    \item The procedure 1--3 is repeated five times with non-intersecting validation datasets so as to produce a validation prediction for all the data that is available, and the accuracy of each validation is stored. 
    \item The accuracy of predictions on each of the five validation datasets are combined to deliver an overall coefficient of determination ($R^2$ value) calculated over the entire dataset.  
\end{enumerate}
Throughout the following section the $R^2$ value calculated in this manner has been used as the metric to compare the different GPR models. We have also introduced the distribution of uncertainty quality (DUQ) described in section~\ref{subsubsect:DUQ} as a metric to further explore the quality of a model's predictions. 
Training on the full database took two minutes on a laptop with an Intel Core i7 CPU---once trained predictions can be made effectively instantaneously. 

The Gaussian RBF (Eq.~\ref{eq:gaussian_rbf_kernel}), linear (Eq.~\ref{eq:linear_kernel}), and ARD kernels were initially tested for the composition only, $k(\mathbf{x},\mathbf{x'})=k(\mathbf{x}_\mathrm{c},\mathbf{x'}_\mathrm{c})$. The results for selected properties are shown in Table~\ref{tab:kernel_results}. 
Using the linear kernel is broadly the same as the Alloy Design Program~\cite{Harada1988PhaseSuperalloys,Harada1999DesignSuperalloys,Kawagishi2012DevelopmentTMS-238,Sato2008AProcessability,Enomoto1991CalculationMethod,Saito1997TheMetals}, which gave good results for the $\gamma'$ fraction, but was less accurate for some elements in the $\gamma$ phase. The RBF kernel gave slight improvements for Ni and Al in both phases as well as the $\gamma'$ fraction. The ARD kernel performed slightly worse than the standard RBF kernel overall, this is due to overfitting, as has been noted in previous work~\cite{Mohammed2017Over-fittingRegression}. In the case of Gaussian RBF kernels, it was found overfitting could be reduced by setting the minimum value of the length scale to $l=0.3$, corresponding to about half the phase diagram~\cite{Duvenaud2014AutomaticProcesses}.

\begin{table}[h]
\centering
\begin{tabular}{lccccc}
\hline
\multicolumn{1}{c}{Kernel}                      & \multicolumn{2}{c}{$\gamma$ phase elements}     & \multicolumn{2}{c}{$\gamma'$ phase elements}    & \multicolumn{1}{c}{$\gamma'$} \\
\multicolumn{1}{c}{$k(\mathbf{x},\mathbf{x}')$} & \multicolumn{1}{c}{Ni} & \multicolumn{1}{c}{Al} & \multicolumn{1}{c}{Ni} & \multicolumn{1}{c}{Al} & \multicolumn{1}{c}{frac.}  \\ \hline \hline
Linear                                            & 0.921                  & 0.040                   & 0.669                  & 0.591                   & 0.900                          \\
RBF                                               & 0.940                  & 0.371                   & 0.704                  & 0.676                   & 0.905                          \\
ARD                                               & 0.936                  & 0.349                   & 0.724                  & 0.658                   & 0.896                          \\ \hline
Optimal (Eq. \ref{eq:good_kernel_2})                                  & 0.927                  & 0.599                   & 0.824                  & 0.674                   & 0.917                          \\ \hline
\end{tabular}
\caption{Comparison of coefficients of determination ($R^2$) for different kernels, including the optimal final kernel. Refer to text for explicit form of kernels.}
\label{tab:kernel_results}
\end{table}

With both linear and RBF kernels working well, a systematic approach was then taken to combining them according to the kernel scheme given in Eq.~\ref{eq:kernel_scheme_2}. 
The best results were found for the following kernel, henceforth referred to as the optimal kernel:
\begin{align}
    k(\mathbf{x},\mathbf{x}')=
        b+a_0\ \mathbf{x}_\mathrm{c}\cdot{\mathbf{x}_\mathrm{c}}'
    +
    a_1\mathrm{e}^{-{\|B(\mathbf{x}_\mathrm{c}-{\mathbf{x}_\mathrm{c}}')\|}^2/l_1^2} \cdot \mathrm{e}^{-{\|\mathbf{x}_\mathrm{h}-{\mathbf{x}_\mathrm{h}}'\|}^2/l_2^2}
    \label{eq:good_kernel_2} \quad . 
\end{align}
The first part is a linear kernel and the second an RBF kernel that combines composition and heat treatment, 
$B$ is a low rank matrix found from principal component analysis~\cite{Duvenaud2014AutomaticProcesses}. A projection onto the first 3 principal components was found to be optimal, in agreement with the discussion in section \ref{subsect:ml}. 

\begin{figure}
    \centering
    \includegraphics[width=\columnwidth]{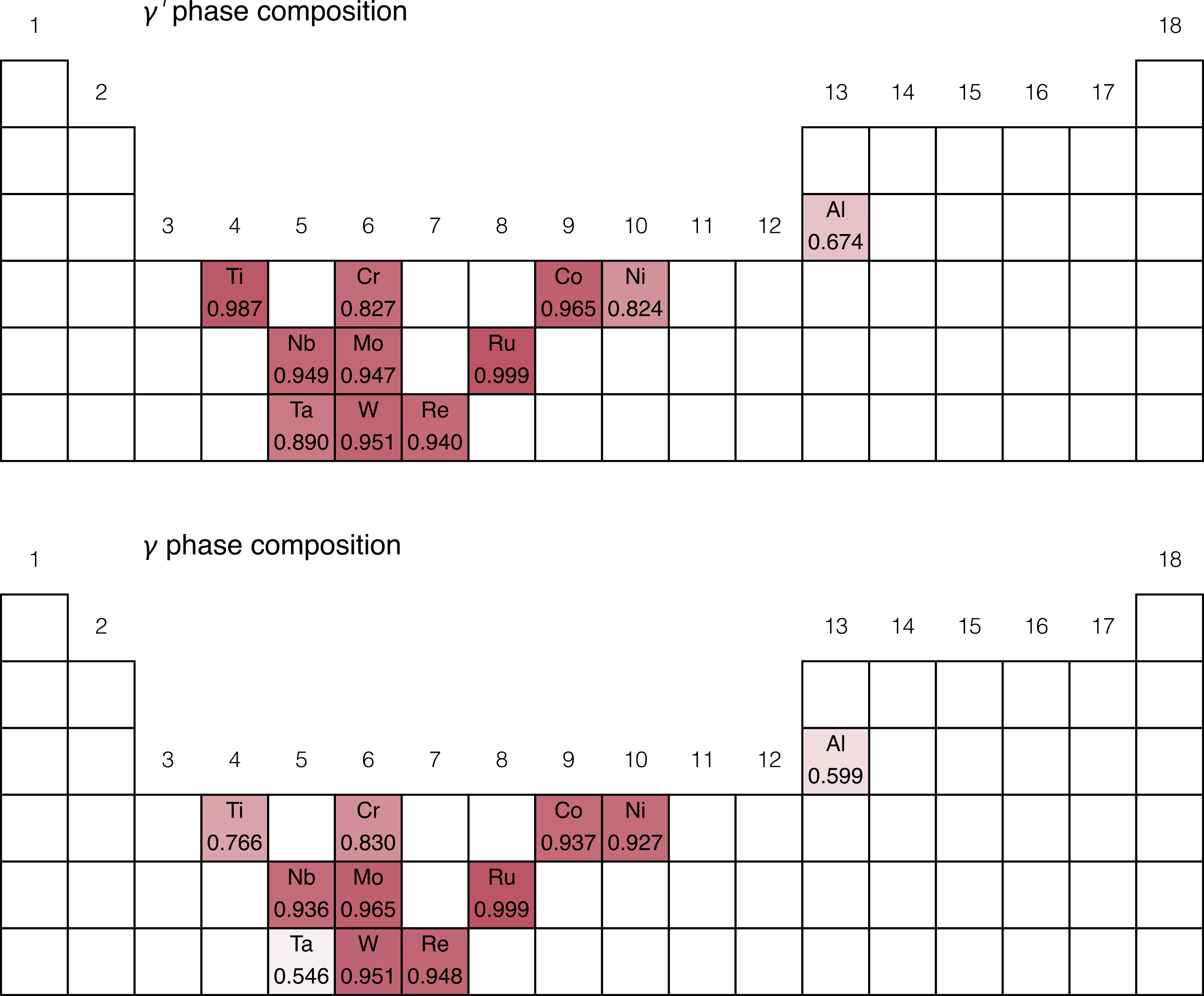}
    \caption{Overview of $R^2$ values obtained for predictions of phase composition using the final GPR model. Elements have their respective $R^2$ listed below them and have also been colour coded to reflect this value (darker is better). The data has been superimposed on the periodic table to reflect how the results correlate to element groupings (group numbers are at the top of each column).}
    \label{fig:final_model_overview}
\end{figure}

Kernel Eq.~\ref{eq:good_kernel_2} gave excellent results for the $\gamma'$ fraction with $R^2=0.917$, and for all components $R^2>0.8$ (the full set of $R^2$ values for this model are given in Fig.~\ref{fig:final_model_overview}) except Al in the $\gamma'$ phase ($R^2=0.674$) and Al, Ta and Ti in the $\gamma$ phase ($R^2=0.599$, $0.546$, $0.766$ respectively). 
This is due to the role of these three elements as $\gamma'$ formers. This role leads to their at. \% in the $\gamma'$ phase being more constant between different alloys, resulting in a smaller variance and hence lower $R^2$, most notably for Al. Analysis of the root mean-squared error (RMSE) showed it to be comparable to the other components. If Al, Ta, and Ti concentrations in the $\gamma'$ phase have a strong physical correlation to the alloy composition, this is not the case for the $\gamma$ phase content---the remaining amount of each element that does not form the $\gamma'$ phase is dissolved into the $\gamma$ phase---and for this reason it is less strongly correlated to the overall alloy composition which results in the generally lower $R^2$ values for the $\gamma'$ forming elements in the $\gamma$ phase (Table~\ref{tab:kernel_results}). 

The predicted $\gamma'$ fractions found from this model are compared to the experimental values in 
Fig.~\ref{fig:frac_final}. 
Agreement is on the whole excellent. The most significant outlier is lab alloy Ni-9.1Al-4.4Ti. This alloy is peculiar in having a low number of components with similar at. \% values across both phase compositions and the nominal composition, which leads to a large error and uncertainty according to eqs.~\ref{eq:calc_f}--\ref{eq:calc_f_std}~\cite{Ralph1982TheTechniques}. Two other outliers are similar to training set alloys but with the addition of Re, suggesting more data for Re-bearing alloys may be required to explain its anomalous effect~\cite{Segersall2015Thermal-mechanicalAlloying,Yoon2007EffectsObservations}. Finally, two outliers correspond to the commercial single-crystal superalloy MC2 with extreme 3rd stage heat treatments (1050$^\circ$C  for 10 hours and 100 hours respectively)~\cite{Duval1994PhaseMC2}---again, more data is needed for extreme outlying heat treatment regimes, especially in the 3rd stage. 

A variant of the optimal kernel Eq.~\ref{eq:good_kernel_2} was also tested without the heat treatment term of the kernel which was set to 1. It performed similarly to the optimal kernel for phase composition predictions but was significantly outperformed for predictions of the $\gamma'$ phase fraction: it achieved an $R^2=0.767$ compared to $0.917$ for the optimal kernel. This tells us the optimal model is capturing the variance due to heat treatment, even if it is not capturing the full physics of complex heat treatment regimes such as the case of MC2 described above. 

Fig.~\ref{fig:Al_in_detail} 
shows the predictions for one crucial component---the Al content of the $\gamma'$ phase. Al is the principal former of the $\gamma'$ phase and the primary constituent of the secondary sublattice in this phase~\cite{Reed2006TheApplications,Durand-Charre1997TheSuperalloys}. Significant outliers are highlighted in the figure. Of these UMF20 and UMF31 are unusual compared to the other alloys because Ti rather than Al is the primary $\gamma'$ phase forming element in their composition~\cite{Delargy1983PhaseIN939,Ma2007DevelopmentSuperalloys,Wlodek1996TheDT}. In the case of STAL-15 and STAL-15-Re, they have similar input compositions whilst their reported phase compositions differ greatly hinting that they are either near to a phase transition or a possible anomalous result~\cite{Segersall2015Thermal-mechanicalAlloying}. 

\begin{figure}
    \centering
    \includegraphics[width=4.8in]{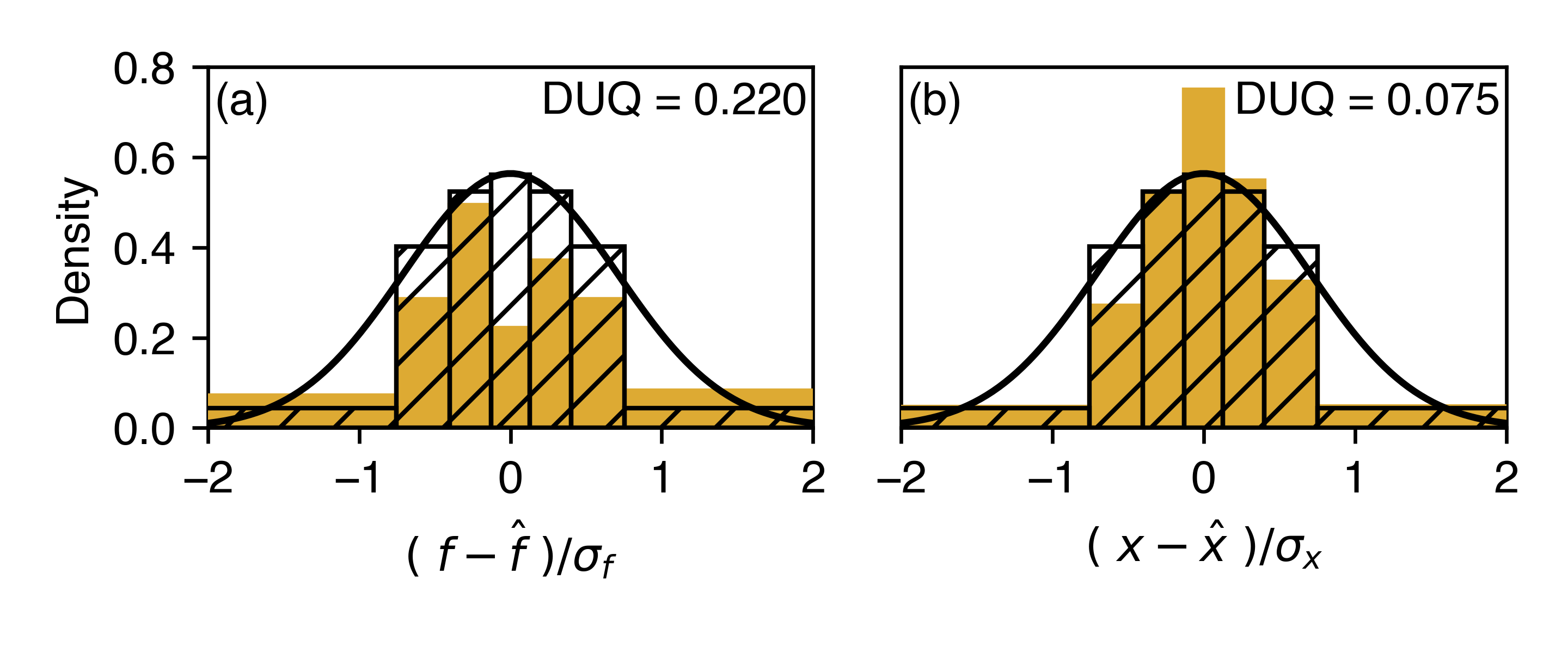}
    \caption{Histograms for the uncertainties associated with predictions. The yellow histogram is for the data and the hatched histogram is for an ideal normal distribution (solid bold line) with the same bins. Left (a): histogram for precipitate fraction data. Right (b): Histogram for all the chemical composition data (all elements across both phases).}
    \label{fig:histograms}
\end{figure}

A key feature of GPR is the ability to estimate uncertainty in its predictions. Therefore, in Fig.~\ref{fig:histograms} we compare the error in a prediction measured against the experimental value to the uncertainty in the prediction, which should take a single value. 
We study both the precipitate fraction and the chemical compositions. The area under each histogram in normalised to 1. For compositions the distribution is symmetrical and similar to the Gaussian distribution, with a comparatively small $\mathrm{DUQ}=0.075$ (see Eq.~\ref{eq:uncertainty_quality}) . For the precipitate fraction $f$ the DUQ is larger at $0.220$, and the distribution has a skew to under/overestimation. 

\begin{figure}[ht]
    \centering
    \includegraphics[width=4in]{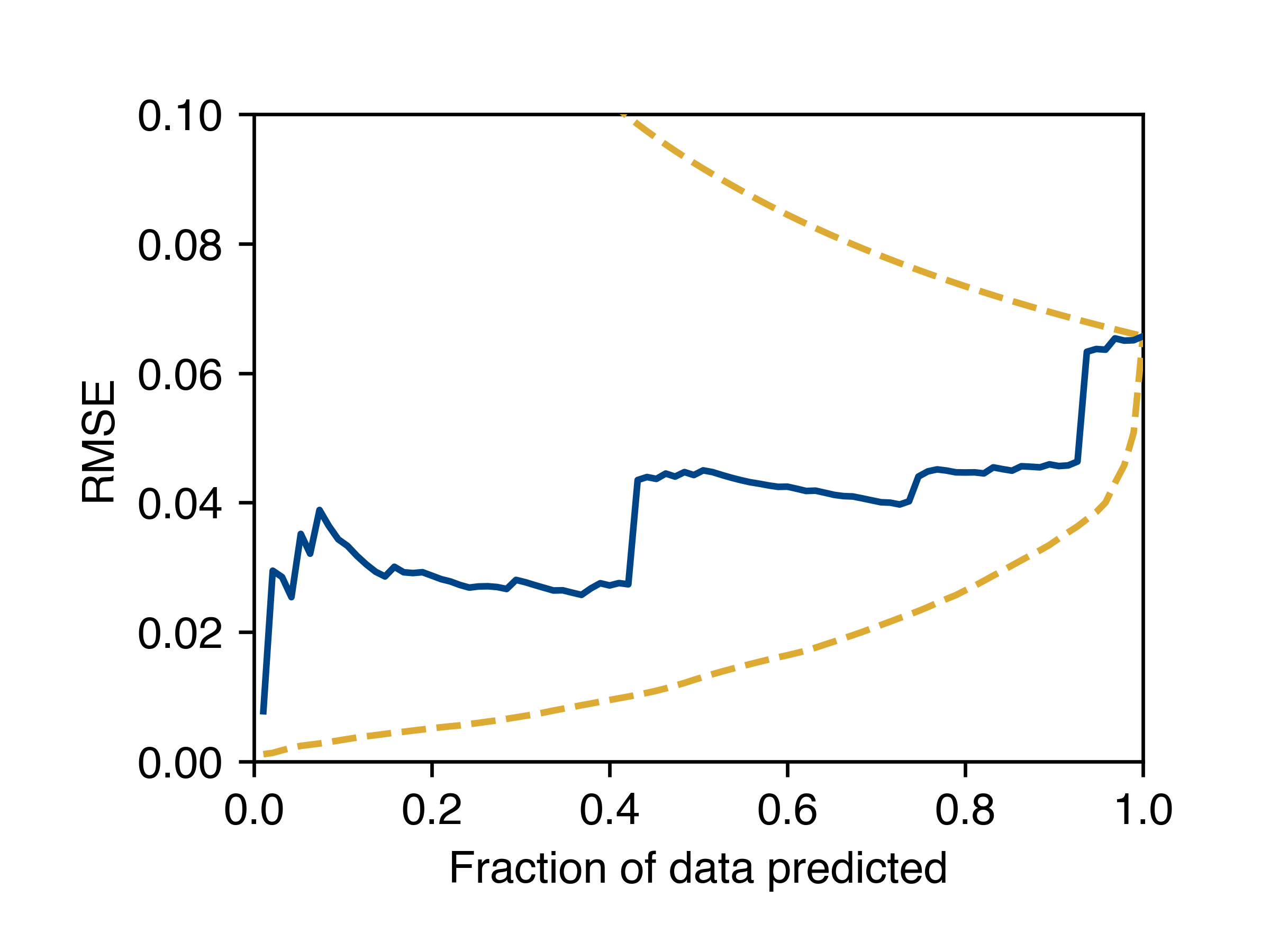}
    \caption{Plot of the calculated RMSE when predictions with the highest uncertainty are filtered out (blue line). The yellow dash-dotted lines give theoretical upper and lower bounds.
    }
    \label{fig:MSFE_with_data_removed}
\end{figure}

A practical advantage of having quantified uncertainties is that predictions can be filtered based on this uncertainty~\cite{Martin2017Profile-QSARCompounds,Irwin2020PracticalData}. Fig.~\ref{fig:MSFE_with_data_removed} shows an example of this for the precipitate fraction. By focusing on only the most certain results, the RMSE reduces significantly, and is closer to the theoretical lower bound for the reduction than the upper bound. The lower and upper bound are given by filtering the data in the order that minimises or maximises respectively the RMSE. 
The steps in the blue curve in Fig.~\ref{fig:MSFE_with_data_removed} occur as significant outlying predictions are filtered out. 
Each step corresponds to a significant outlier that can be identified on Fig.~\ref{fig:frac_final} by the magnitude of its uncertainty; for example the step at the highest fraction of data predicted corresponds to Ni-8.5Cr-10Al-2Re, the next MC2, then Ni-9.1Al-4.4Ti, etc. 
Step-like behaviour would vanish if all the predictions with the largest errors had the largest uncertainties, i.e.\ the lower bound curve in yellow. 
This method is useful when the model's predictions are to be used to choose new alloy compositions for experimental testing, as it allows the alloy designer to only test compositions below a certain predicted error threshold and 
to focus on the predictions most likely to fulfil the target tolerance in an experiment.

\subsection{Comparison to CALPHAD}
The final GPR model was trained on the full database with 93 entries. It was then tested against unseen experimental atom probe tomography data collected by Sulzer \textit{et al} for the SX-series alloys specifically for benchmarking the performance of CALPHAD~\cite{Sulzer2020TheApplications} (data highlighted with an asterisk in \ref{sect:data_appendix}). 
In that test two thermodynamic databases were tested, TTNi8 and TCNi8~\cite{Andersson2002Thermo-CalcScience}. 
For each prediction tool, the root mean-squared error (RMSE) was calculated over all the elements in each phase in each SX-series alloy, as well as the precipitate fraction $f$, giving three metrics for each method (Table \ref{tab:calphad_comp_summary}). For both the $\gamma$ and $\gamma'$ phases and the precipitate fraction $f$, the GPR model is more accurate than CALPHAD. 

\begin{table}[h]
\centering
\begin{tabular}{lrrr}
\hline
\multicolumn{1}{c}{} & \multicolumn{1}{c}{$\gamma'$ phase} & \multicolumn{1}{c}{$\gamma$ phase} & \multicolumn{1}{c}{$\gamma'$ fraction $f$} \\
\multicolumn{1}{c}{Method} & \multicolumn{1}{c}{(RMSE)} & \multicolumn{1}{c}{(RMSE)} & \multicolumn{1}{c}{(RMSE)} \\ \hline \hline
TCNi8 & 0.0106 & 0.0354 & 0.0371 \\
TTNi8 & 0.0072 & 0.0324 & 0.0205 \\
This work & 0.0065 & 0.0321 & 0.0201 \\ \hline
\end{tabular}
\caption{Root mean-squared error (RMSE) for the predictions of the GPR model of this work and CALPHAD models on the SX-series superalloys using the TCNi8 and TTNi8 databases.}
\label{tab:calphad_comp_summary}
\end{table}

The composition of the SX series alloys in Ref.~\cite{Sulzer2020TheApplications} had been chosen to explore a new region of composition space with high Cr content, and consequently both the GPR model and the two CALPHAD models were extrapolating in this region. 
Due to the inherent inclusion of physical laws it was anticipated that CALPHAD would be superior for extrapolative predictions, so it is particularly impressive that the GPR model outperforms CALPHAD. 
A further advantage of the GPR method is that it quantifies the uncertainty associated with predictions.

\section{Conclusion}
A Gaussian process regression model was developed to predict the compositional microstructure of nickel-base superalloys. It inputs the nominal composition of a superalloy and the ageing heat treatments, to then predict the chemical composition of the key $\gamma$ and $\gamma'$ phases, and their relative abundance. The composition predictions are combined via a probabilistic approach to produce the final output composition. 
Cross-validation was used to compare a number of different kernel schemes for the GPR models, with the optimal kernel achieving a coefficient of determination $R^2=0.917$ for the precipitate fraction. 

The real life utility of the GPR model was demonstrated by predicting the composition of the benchmark SX-series superalloys. The GPR model outperformed CALPHAD predictions using both the TTNi8 and TCNi8 databases~\cite{Sulzer2020TheApplications}. 
This demonstrates the benefits of combining the best of physical and statistical approaches.
The GPR model has a number of additional advantages over CALPHAD:
\begin{itemize}
    \item Returns quantified uncertainty estimates for each prediction.
    \item No prior thermodynamic knowledge is required to construct the model.
    \item Model can easily be retrained as more data becomes available. 
    \item Additional non-equilibrium effects can be incorporated in the model.
    \item Once trained, no free energy minimisation step is required to make predictions from the model. 
    \item The trained model can be used to identify outliers in the initial dataset by highlighting property entries that lie the furthest from the model's predictions~\cite{Verpoort2018MaterialsNetwork}.
\end{itemize}

As discussed above our GPR model made better predictions when the kernel included a heat treatment component, however further work is required to determine whether it can precisely capture the evolution of an alloy's microstructure with different applied ageing heat treatments. Training and assessing the model on this problem is difficult due to the paucity of experimental data where phase composition has been determined for superalloys with the same nominal composition but different ageing heat treatments~\cite{Duval1994PhaseMC2,Schmidt1992Effect99}. A better representation of the heat treatment descriptors could improve this. 
Thermodynamic modelling approaches to predicting the effects of heat treatments do exist but their use in optimising heat treatments for alloys of pre-determined composition is involved~\cite{Collins2013GrainSuperalloy,Collins2014ASuperalloy}. 

In this work we opted for the most straightforward dynamic method that makes use of the uncertainty in the initial predictions, Eq.~\ref{eq:comp_adjustment_choice}. We found that this gave improved results for crucial elements such as Ni and Al compared to a fixed balance element model (Section \ref{subsubsect:bal_element}). Alternate dynamic schemes for balancing the total phase composition in order to fulfil Eq.~\ref{eq:comp_constraint} can be devised, which may produce further improvements. 

The GPR model is completely generic and can be applied to other material systems, meaning that it can be extended to predict other properties as well as thermodynamic ones. This would enable the development of a complete machine learning tool able to design practical alloys that simultaneously satisfy a range of target properties. 

\section*{CRediT authorship contribution statement}
\textbf{Patrick Taylor}: Conceptualisation, Methodology, Software, Formal analysis, Data curation, Writing---original draft. \textbf{Gareth Conduit}: Conceptualisation, Methodology, Writing---review \& editing, Supervision.

\section*{Declaration of competing interest}
Gareth Conduit is a Director of materials machine learning company Intellegens. The authors have no other financial interests or personal relationships that could have appeared to influence the work reported in this paper.

\section*{Acknowledgements}
Patrick Taylor acknowledges the financial support of EPSRC and an ICASE award from Dassault Syst\`emes UK. Gareth Conduit acknowledges the financial support of the Royal Society. We would also like to thank Dr.\ Victor Milman and Dr.\ Alexander Perlov at Dassault Syst\`emes UK for their feedback on the manuscript. 
There is Open Access to this paper and data available at https://www.openaccess.cam.ac.uk/.

\bibliographystyle{elsarticle-num}
\bibliography{references}

\newpage
\appendix
\setcounter{table}{0}
\setcounter{equation}{0}
\renewcommand{\thetable}{A\arabic{table}}
\renewcommand{\theequation}{A\arabic{equation}}

\section{Distribution of uncertainty quality}
The distribution of uncertainty quality (DUQ) as defined in Eq.~\ref{eq:uncertainty_quality} has a minimum value of 0 and a maximum value of 1 which can be seen by considering the $\ell_1$ norm on the vectors $\mathbf{n}=[n_1,...,n_m,...,n_M]^\mathrm{T}$. For both the ideal and real histogram the edges of each bin $z_m$ are defined such that:
\begin{align}
    \int_{z_{m-1}}^{z_m} \varphi(z) \mathrm{d}z = A \quad ,
\end{align}
where $\varphi(z)$ is the PDF for the Gaussian distribution $\mathcal{N}(0,1)$, and the area of each ideal bin is $A=1/M$. This ensures the histogram for the ideal distribution has a total area of 1. 
For an odd number of bins the rightmost edges $z_m$ with $z_m>0$ are given by:
\begin{align}
    z_m = \mathrm{erf}^{-1}\left((2m-1)\frac{1}{M}\right) \quad , \label{eq:GUQ_bins}
\end{align}
where $m=1,2,...,(M+1)/2$. 

The number of bins is optimised by minimising a quantity related to the DUQ:
\begin{align}
    \frac{1}{2}\frac{M}{M-1}\sum_{m}^{M}\left\lvert \frac{n_m}{M}-\frac{1}{M}\right\rvert 
    + \int_\infty^\infty \left\lvert h(z) - \varphi(z) \right\rvert \mathrm{d}z \quad ,
\end{align}
where $h(z)$ is the height of the ideal bin at position $z$. The second term accounts for the difference in area between the ideal bins and the normal distribution, which favours larger numbers of bins. Minimising the DUQ directly will always lead to the minimum number of bins being used. 

Eq.~\ref{eq:GUQ_bins} leads to the outermost bins having infinite width and zero height. For visualisation purposes this is undesirable so a cutoff can be introduced that gives all the bins finite width whilst not affecting the optimal number of bins. We found a cutoff $z_\mathrm{c}=\pm\ \mathrm{erf}^{-1}(0.999999)=\pm 3.459$ to be a suitable choice. 

\begin{landscape}
\section{Database of alloys}
\label{sect:data_appendix}
The full database of alloys as described in section~\ref{sect:database_processing}. The SX-series alloys have been highlighted in each table with an asterisk. 
\\
\begin{center}
\small
\begin{longtable}{llrrrrrrrrrrrr}
\hline
\multicolumn{1}{c}{\#} & \multicolumn{1}{c}{Ref.} & \multicolumn{12}{c}{Composition at. \%} \\
\multicolumn{1}{c}{} & \multicolumn{1}{c}{} & \multicolumn{1}{c}{Ni} & \multicolumn{1}{c}{Cr} & \multicolumn{1}{c}{Co} & \multicolumn{1}{c}{Re} & \multicolumn{1}{c}{Ru} & \multicolumn{1}{c}{Al} & \multicolumn{1}{c}{Ta} & \multicolumn{1}{c}{W} & \multicolumn{1}{c}{Ti} & \multicolumn{1}{c}{Mo} & \multicolumn{1}{c}{Nb} & \multicolumn{1}{c}{Hf} \\ \hline\hline
1 & \cite{Duval1994PhaseMC2} & 65.90 & 9.30 & 5.10 & - & - & 11.20 & 2.00 & 2.60 & 2.50 & 1.30 & - & - \\
2 & \cite{Duval1994PhaseMC2} & 65.90 & 9.30 & 5.10 & - & - & 11.20 & 2.00 & 2.60 & 2.50 & 1.30 & - & - \\
3 & \cite{Duval1994PhaseMC2} & 65.90 & 9.30 & 5.10 & - & - & 11.20 & 2.00 & 2.60 & 2.50 & 1.30 & - & - \\
4 & \cite{Duval1994PhaseMC2} & 65.90 & 9.30 & 5.10 & - & - & 11.20 & 2.00 & 2.60 & 2.50 & 1.30 & - & - \\
5 & \cite{Glas1996OrderSuperalloys} & 66.50 & 8.87 & 5.38 & - & - & 12.81 & 1.11 & 1.56 & 2.41 & 1.35 & - & - \\
6 & \cite{Harada1993Atom-probeSuperalloy} & 72.00 & 7.80 & - & - & - & 12.70 & 2.80 & - & - & 4.60 & - & - \\
7 & \cite{Harada1988PhaseSuperalloys} & 65.30 & 6.75 & 8.12 & - & - & 12.29 & 1.80 & 5.76 & - & - & - & - \\
8 & \cite{Harada1988PhaseSuperalloys} & 64.66 & 11.68 & 5.15 & - & - & 11.25 & 4.03 & 1.32 & 1.90 & - & - & - \\
9 & \cite{Harada1988PhaseSuperalloys} & 66.29 & 9.22 & 5.09 & - & - & 13.55 & 1.99 & 2.61 & - & 1.25 & - & - \\
10 & \cite{Harada1988PhaseSuperalloys,Jalilvand2013InfluenceIN-738LC} & 59.89 & 17.67 & 8.23 & - & - & 7.19 & 0.54 & 0.81 & 4.05 & 1.07 & 0.55 & - \\
11 & \cite{Harada1988PhaseSuperalloys,Collier1986EffectsProperties.} & 62.81 & 9.19 & 9.72 & - & - & 12.33 & 1.25 & 0.03 & 1.15 & 3.51 & 0.01 & - \\
12 & \cite{Harada1988PhaseSuperalloys,Basak2017MicrostructureSLE} & 61.22 & 9.17 & 10.11 & - & - & 13.25 & 0.99 & 3.24 & 1.24 & 0.43 & - & 0.33 \\
13 & \cite{Harada1988PhaseSuperalloys} & 67.77 & 9.19 & 4.66 & - & - & 12.40 & 1.91 & 2.57 & 1.12 & 0.37 & - & - \\
14 & \cite{Khan1986EffectCMSX-2} & 67.77 & 9.19 & 4.66 & - & - & 12.40 & 1.91 & 2.57 & 1.12 & 0.37 & - & - \\
15 & \cite{Khan1986EffectCMSX-2} & 67.77 & 9.19 & 4.66 & - & - & 12.40 & 1.91 & 2.57 & 1.12 & 0.37 & - & - \\
16 & \cite{Khan1984TheSuperalloy} & 71.38 & 9.22 & - & - & - & 13.55 & 1.99 & 2.61 & - & 1.25 & - & - \\
17 & \cite{Khan1984TheSuperalloy} & 66.29 & 9.22 & 5.09 & - & - & 13.55 & 1.99 & 2.61 & - & 1.25 & - & - \\
18 & \cite{Khan1984TheSuperalloy} & 63.74 & 9.22 & 7.63 & - & - & 13.56 & 1.99 & 2.61 & - & 1.25 & - & - \\
19 & \cite{Miller1994APFIMSuperalloy} & 65.10 & 11.50 & 5.10 & - & - & 11.00 & 4.00 & 1.40 & 1.90 & - & - & - \\
20 & \cite{Miller1994APFIMSuperalloy} & 65.10 & 11.50 & 5.10 & - & - & 11.00 & 4.00 & 1.40 & 1.90 & - & - & - \\
21 & \cite{Royer1998InSuperalloy} & 65.60 & 8.70 & 6.60 & - & - & 11.80 & 2.70 & 1.80 & 1.50 & 1.30 & - & - \\
22 & \cite{Diologent2004OnSuperalloys} & 65.54 & 9.04 & 6.65 & - & - & 11.62 & 2.63 & 1.87 & 1.39 & 1.26 & - & - \\
23 & \cite{Diologent2004OnSuperalloys} & 65.54 & 9.04 & 6.65 & - & - & 11.62 & 2.63 & 1.87 & 1.39 & 1.26 & - & - \\
24 & \cite{Diologent2004OnSuperalloys} & 65.54 & 9.04 & 6.65 & - & - & 11.62 & 2.63 & 1.87 & 1.39 & 1.26 & - & - \\
25 & \cite{Segersall2015Thermal-mechanicalAlloying} & 62.85 & 17.74 & 5.06 & - & - & 9.94 & 2.56 & 1.23 & - & 0.62 & - & 0.03 \\
26 & \cite{Segersall2015Thermal-mechanicalAlloying} & 62.52 & 17.81 & 5.06 & 0.64 & - & 10.03 & 2.64 & 0.65 & - & 0.61 & - & 0.04 \\
27 & \cite{Schmidt1992Effect99} & 66.70 & 9.60 & 5.00 & - & - & 12.00 & 0.90 & 3.00 & 2.70 & - & - & - \\
28 & \cite{Schmidt1992Effect99} & 66.70 & 9.60 & 5.00 & - & - & 12.00 & 0.90 & 3.00 & 2.70 & - & - & - \\
29 & \cite{Schmidt1992Effect99} & 66.70 & 9.60 & 5.00 & - & - & 12.00 & 0.90 & 3.00 & 2.70 & - & - & - \\
30 & \cite{Schmidt1992Effect99} & 66.70 & 9.60 & 5.00 & - & - & 12.00 & 0.90 & 3.00 & 2.70 & - & - & - \\
31 & \cite{Loomis1972TheSuperalloys} & 78.34 & 15.19 & - & - & - & 6.47 & - & - & - & - & - & - \\
32 & \cite{Loomis1972TheSuperalloys} & 77.41 & 14.95 & - & - & - & 6.44 & - & - & - & 1.20 & - & - \\
33 & \cite{Loomis1972TheSuperalloys} & 75.79 & 14.60 & - & - & - & 6.50 & - & - & - & 3.11 & - & - \\
34 & \cite{Loomis1972TheSuperalloys} & 73.89 & 14.24 & - & - & - & 6.71 & - & - & - & 5.16 & - & - \\
35 & \cite{Loomis1972TheSuperalloys} & 76.06 & 15.06 & - & - & - & 8.88 & - & - & - & - & - & - \\
36 & \cite{Loomis1972TheSuperalloys} & 75.16 & 14.81 & - & - & - & 8.85 & - & - & - & 1.18 & - & - \\
37 & \cite{Loomis1972TheSuperalloys} & 73.64 & 14.43 & - & - & - & 8.96 & - & - & - & 2.97 & - & - \\
38 & \cite{Loomis1972TheSuperalloys} & 72.08 & 14.05 & - & - & - & 9.07 & - & - & - & 4.80 & - & - \\
39 & \cite{Loomis1972TheSuperalloys} & 74.34 & 14.05 & - & - & - & 11.61 & - & - & - & - & - & - \\
40 & \cite{Loomis1972TheSuperalloys} & 73.15 & 13.94 & - & - & - & 11.76 & - & - & - & 1.15 & - & - \\
41 & \cite{Loomis1972TheSuperalloys} & 71.45 & 13.59 & - & - & - & 12.12 & - & - & - & 2.84 & - & - \\
42 & \cite{Loomis1972TheSuperalloys} & 70.03 & 13.42 & - & - & - & 12.09 & - & - & - & 4.46 & - & - \\
43 & \cite{Loomis1972TheSuperalloys} & 78.74 & 15.14 & - & - & - & 2.00 & - & - & 4.12 & - & - & - \\
44 & \cite{Loomis1972TheSuperalloys} & 77.80 & 14.93 & - & - & - & 2.00 & - & - & 4.08 & 1.19 & - & - \\
45 & \cite{Loomis1972TheSuperalloys} & 76.16 & 14.59 & - & - & - & 2.20 & - & - & 4.09 & 2.96 & - & - \\
46 & \cite{Loomis1972TheSuperalloys} & 74.67 & 14.25 & - & - & - & 2.10 & - & - & 4.14 & 4.84 & - & - \\
47 & \cite{Loomis1972TheSuperalloys} & 76.06 & 15.06 & - & - & - & 8.88 & - & - & - & - & - & - \\
48 & \cite{Loomis1972TheSuperalloys} & 75.16 & 14.81 & - & - & - & 8.85 & - & - & - & 1.18 & - & - \\
49 & \cite{Loomis1972TheSuperalloys} & 73.64 & 14.43 & - & - & - & 8.96 & - & - & - & 2.97 & - & - \\
50 & \cite{Loomis1972TheSuperalloys} & 72.08 & 14.05 & - & - & - & 9.07 & - & - & - & 4.80 & - & - \\
51 & \cite{Loomis1972TheSuperalloys} & 74.34 & 14.05 & - & - & - & 11.61 & - & - & - & - & - & - \\
52 & \cite{Loomis1972TheSuperalloys} & 73.15 & 13.94 & - & - & - & 11.76 & - & - & - & 1.15 & - & - \\
53 & \cite{Loomis1972TheSuperalloys} & 71.45 & 13.59 & - & - & - & 12.12 & - & - & - & 2.84 & - & - \\
54 & \cite{Loomis1972TheSuperalloys} & 70.03 & 13.42 & - & - & - & 12.09 & - & - & - & 4.46 & - & - \\
55 & \cite{Loomis1972TheSuperalloys} & 78.74 & 15.14 & - & - & - & 2.00 & - & - & 4.12 & - & - & - \\
56 & \cite{Loomis1972TheSuperalloys} & 77.80 & 14.93 & - & - & - & 2.00 & - & - & 4.08 & 1.19 & - & - \\
57 & \cite{Loomis1972TheSuperalloys} & 76.16 & 14.59 & - & - & - & 2.20 & - & - & 4.09 & 2.96 & - & - \\
58 & \cite{Loomis1972TheSuperalloys} & 74.67 & 14.25 & - & - & - & 2.10 & - & - & 4.14 & 4.84 & - & - \\
59 & \cite{Blavette1986AnSuperalloys} & 64.66 & 11.68 & 5.15 & - & - & 11.26 & 4.03 & 1.32 & 1.90 & - & - & - \\
60 & \cite{Blavette1986AnSuperalloys} & 64.67 & 11.68 & 5.15 & 1.31 & - & 11.26 & 4.03 & - & 1.90 & - & - & - \\
61 & \cite{Blavette1986AnSuperalloys} & 67.33 & 9.22 & 5.08 & - & - & 12.21 & 1.99 & 2.61 & 1.25 & 0.31 & - & - \\
62 & \cite{Blavette1986AnSuperalloys} & 67.35 & 9.22 & 5.08 & 1.29 & - & 12.21 & 1.99 & 1.30 & 1.25 & 0.31 & - & - \\
63 & \cite{Reed2004IdentificationTomography,Reed2004IdentificationTomographyb} & 62.94 & 3.00 & 12.80 & 2.20 & - & 14.00 & 1.90 & 3.10 & - & - & - & 0.06 \\
64 & \cite{Reed2004IdentificationTomography,Reed2004IdentificationTomographyb} & 61.24 & 3.00 & 13.00 & 2.20 & 1.30 & 14.20 & 1.90 & 3.10 & - & - & - & 0.06 \\
65 & \cite{Delargy1983PhaseIN939} & 47.28 & 24.47 & 18.23 & - & - & 3.98 & 0.44 & 0.62 & 4.37 & - & 0.61 & - \\
66 & \cite{Ralph1982TheTechniques} & 86.50 & - & - & - & - & 9.10 & - & - & 4.40 & - & - & - \\
67 & \cite{Blavette1996Atomic-scaleSuperalloys} & 52.60 & 15.90 & 16.00 & - & - & 8.50 & - & - & 4.00 & 3.00 & - & - \\
68 & \cite{Yoon2007EffectsObservations} & 79.50 & 8.50 & - & 2.00 & - & 10.00 & - & - & - & - & - & - \\
69 & \cite{Yoon2007EffectsObservations} & 79.50 & 8.50 & - & 2.00 & - & 10.00 & - & - & - & - & - & - \\
70 & \cite{Yoon2007EffectsObservations} & 79.50 & 8.50 & - & 2.00 & - & 10.00 & - & - & - & - & - & - \\
71 & \cite{Yoon2007EffectsObservations} & 79.50 & 8.50 & - & 2.00 & - & 10.00 & - & - & - & - & - & - \\
72 & \cite{Yoon2007EffectsObservations} & 79.50 & 8.50 & - & 2.00 & - & 10.00 & - & - & - & - & - & - \\
73 & \cite{Yoon2007EffectsObservations} & 79.50 & 8.50 & - & 2.00 & - & 10.00 & - & - & - & - & - & - \\
74 & \cite{Yoon2007EffectsObservations} & 79.50 & 8.50 & - & 2.00 & - & 10.00 & - & - & - & - & - & - \\
75 & \cite{Parsa2015AdvancedSuperalloys} & 63.19 & 7.26 & 9.61 & 0.92 & - & 12.86 & 2.29 & 2.19 & 1.27 & 0.38 & - & 0.03 \\
76 & \cite{Ma2007DevelopmentSuperalloys} & 67.00 & 8.00 & 2.50 & 1.50 & 3.50 & 13.90 & 2.80 & 1.00 & - & - & - & - \\
77 & \cite{Ma2007DevelopmentSuperalloys} & 59.00 & 8.00 & 10.50 & 1.50 & 3.50 & 13.80 & 2.70 & 1.00 & - & - & - & - \\
78 & \cite{Ma2007DevelopmentSuperalloys} & 65.50 & 8.00 & 2.50 & 2.50 & 3.50 & 13.70 & 2.80 & 1.50 & - & - & - & - \\
79 & \cite{Ma2007DevelopmentSuperalloys} & 57.50 & 8.00 & 10.50 & 2.50 & 3.50 & 13.70 & 2.80 & 1.50 & - & - & - & - \\
80 & \cite{Collier1986EffectsProperties.} & 53.05 & 19.39 & 14.14 & - & - & 5.43 & - & 0.46 & 5.76 & 1.76 & 0.01 & - \\
81 & \cite{Collier1986EffectsProperties.} & 52.68 & 19.57 & 14.23 & - & - & 5.39 & - & - & 5.84 & 2.27 & 0.01 & - \\
82 & \cite{Collier1986EffectsProperties.} & 62.87 & 9.18 & 9.71 & - & - & 12.32 & 1.25 & - & 1.15 & 3.51 & 0.01 & - \\
83 & \cite{Collier1986EffectsProperties.} & 63.32 & 8.71 & 9.77 & - & - & 12.23 & - & - & 1.17 & 4.78 & 0.02 & - \\
84 & \cite{Collier1986EffectsProperties.} & 59.99 & 17.65 & 8.15 & - & - & 7.04 & 0.54 & 0.81 & 4.24 & 1.04 & 0.53 & - \\
85 & \cite{Collier1986EffectsProperties.} & 60.15 & 17.57 & 8.33 & - & - & 7.11 & 0.53 & - & 3.89 & 1.82 & 0.61 & - \\
86 & \cite{Collier1986EffectsProperties.} & 59.93 & 17.61 & 8.15 & - & - & 7.35 & - & 0.78 & 4.05 & 1.56 & 0.57 & - \\
87 & \cite{Collier1986EffectsProperties.} & 60.23 & 17.46 & 8.30 & - & - & 7.17 & 0.53 & 0.79 & 3.98 & 1.54 & 0.01 & - \\
88 & \cite{Long2018MicrostructuralReview} & 63.54 & 8.02 & 9.13 & 0.94 & - & 13.44 & 2.24 & 1.81 & - & 0.88 & - & - \\
89 & \cite{Wlodek1996TheDT} & 56.06 & 17.92 & 12.85 & - & - & 4.53 & - & 1.27 & 4.50 & 2.43 & 0.44 & - \\
90 & \cite{Lapington2018CharacterizationSuperalloys} & 48.50 & 18.90 & 18.50 & - & - & 7.90 & 0.70 & 1.00 & 4.30 & - & - & - \\
91 & \cite{Lapington2018CharacterizationSuperalloys} & 48.50 & 18.60 & 18.60 & - & - & 7.90 & 0.60 & 0.90 & 4.00 & - & 0.40 & - \\
92 & \cite{Lapington2018CharacterizationSuperalloys} & 49.40 & 18.60 & 18.30 & - & - & 7.60 & 0.60 & 0.90 & 3.30 & - & 0.80 & - \\
93 & \cite{Lapington2018CharacterizationSuperalloys} & 48.10 & 18.90 & 18.80 & - & - & 8.10 & 0.60 & 0.90 & 2.90 & - & 1.20 & - \\
94* & \cite{Sulzer2020TheApplications} & 64.20 & 14.20 & 4.99 & - & - & 12.00 & 2.60 & 1.12 & - & 0.92 & - & - \\
95* & \cite{Sulzer2020TheApplications} & 63.60 & 14.30 & 5.02 & - & - & 12.10 & 2.62 & 1.13 & - & 0.93 & - & 0.33 \\
96* & \cite{Sulzer2020TheApplications} & 63.50 & 14.00 & 4.93 & - & - & 10.80 & 1.93 & 1.11 & 2.91 & 0.91 & - & - \\
97* & \cite{Sulzer2020TheApplications} & 59.20 & 14.20 & 9.98 & - & - & 12.00 & 2.60 & 1.12 & - & 0.92 & - & -
\\ \hline
\end{longtable}


\begin{longtable}{llrrrrrr}
\hline
\multicolumn{1}{c}{\#} & \multicolumn{1}{c}{Ref.} & \multicolumn{6}{c}{Precipitation heat   treatment temp.$\lvert$time ($^\circ$C$\lvert$hrs)} \\
 &  & \multicolumn{2}{c}{\#1} & \multicolumn{2}{c}{\#2} & \multicolumn{2}{c}{\#3} \\
 \hline\hline
1 & \cite{Duval1994PhaseMC2} & 1100 & 4 & 850 & 24 & - & - \\
2 & \cite{Duval1994PhaseMC2} & 1100 & 4 & 850 & 24 & 1050 & 0.5 \\
3 & \cite{Duval1994PhaseMC2} & 1100 & 4 & 850 & 24 & 1050 & 10 \\
4 & \cite{Duval1994PhaseMC2} & 1100 & 4 & 850 & 24 & 1050 & 100 \\
5 & \cite{Glas1996OrderSuperalloys} & 1100 & 4 & 850 & 24 & - & - \\
6 & \cite{Harada1993Atom-probeSuperalloy} & 982 & 5 & 870 & 20 & 1040 & 3 \\
7 & \cite{Harada1988PhaseSuperalloys} & 900 & 1500 & - & - & - & - \\
8 & \cite{Harada1988PhaseSuperalloys} & 900 & 1500 & - & - & - & - \\
9 & \cite{Harada1988PhaseSuperalloys} & 900 & 1500 & - & - & - & - \\
10 & \cite{Harada1988PhaseSuperalloys,Jalilvand2013InfluenceIN-738LC} & 900 & 1500 & - & - & - & - \\
11 & \cite{Harada1988PhaseSuperalloys,Collier1986EffectsProperties.} & 900 & 1500 & - & - & - & - \\
12 & \cite{Harada1988PhaseSuperalloys,Basak2017MicrostructureSLE} & 900 & 1500 & - & - & - & - \\
13 & \cite{Harada1988PhaseSuperalloys} & 900 & 1500 & - & - & - & - \\
14 & \cite{Khan1986EffectCMSX-2} & 980 & 5 & 850 & 48 & - & - \\
15 & \cite{Khan1986EffectCMSX-2} & 1050 & 16 & 850 & 48 & - & - \\
16 & \cite{Khan1984TheSuperalloy} & 1100 & 4 & 850 & 24 & - & - \\
17 & \cite{Khan1984TheSuperalloy} & 1100 & 4 & 850 & 24 & - & - \\
18 & \cite{Khan1984TheSuperalloy} & 1100 & 4 & 850 & 24 & - & - \\
19 & \cite{Miller1994APFIMSuperalloy} & 1079 & 4 & - & - & - & - \\
20 & \cite{Miller1994APFIMSuperalloy} & 1079 & 4 & 871 & 32 & - & - \\
21 & \cite{Royer1998InSuperalloy} & 1050 & 16 & - & - & - & - \\
22 & \cite{Diologent2004OnSuperalloys} & 1000 & 2.5 & 850 & 24 & - & - \\
23 & \cite{Diologent2004OnSuperalloys} & 1100 & 9 & 850 & 24 & - & - \\
24 & \cite{Diologent2004OnSuperalloys} & 1180 & 0.25 & 1160 & 8 & 850 & 24 \\
25 & \cite{Segersall2015Thermal-mechanicalAlloying} & 1300 & 5 & 1100 & 6 & 850 & 20 \\
26 & \cite{Segersall2015Thermal-mechanicalAlloying} & 1100 & 6 & 850 & 20 & - & - \\
27 & \cite{Schmidt1992Effect99} & - & - & - & - & - & - \\
28 & \cite{Schmidt1992Effect99} & 870 & 16 & - & - & - & - \\
29 & \cite{Schmidt1992Effect99} & 1080 & 4 & 870 & 16 & - & - \\
30 & \cite{Schmidt1992Effect99} & 1080 & 32 & 870 & 16 & - & - \\
31 & \cite{Loomis1972TheSuperalloys} & 760 & 1000 & - & - & - & - \\
32 & \cite{Loomis1972TheSuperalloys} & 760 & 1000 & - & - & - & - \\
33 & \cite{Loomis1972TheSuperalloys} & 760 & 1000 & - & - & - & - \\
34 & \cite{Loomis1972TheSuperalloys} & 760 & 1000 & - & - & - & - \\
35 & \cite{Loomis1972TheSuperalloys} & 760 & 1000 & - & - & - & - \\
36 & \cite{Loomis1972TheSuperalloys} & 760 & 1000 & - & - & - & - \\
37 & \cite{Loomis1972TheSuperalloys} & 760 & 1000 & - & - & - & - \\
38 & \cite{Loomis1972TheSuperalloys} & 760 & 1000 & - & - & - & - \\
39 & \cite{Loomis1972TheSuperalloys} & 760 & 1000 & - & - & - & - \\
40 & \cite{Loomis1972TheSuperalloys} & 760 & 1000 & - & - & - & - \\
41 & \cite{Loomis1972TheSuperalloys} & 760 & 1000 & - & - & - & - \\
42 & \cite{Loomis1972TheSuperalloys} & 760 & 1000 & - & - & - & - \\
43 & \cite{Loomis1972TheSuperalloys} & 760 & 1000 & - & - & - & - \\
44 & \cite{Loomis1972TheSuperalloys} & 760 & 1000 & - & - & - & - \\
45 & \cite{Loomis1972TheSuperalloys} & 760 & 1000 & - & - & - & - \\
46 & \cite{Loomis1972TheSuperalloys} & 760 & 1000 & - & - & - & - \\
47 & \cite{Loomis1972TheSuperalloys} & 927 & 100 & - & - & - & - \\
48 & \cite{Loomis1972TheSuperalloys} & 927 & 100 & - & - & - & - \\
49 & \cite{Loomis1972TheSuperalloys} & 927 & 100 & - & - & - & - \\
50 & \cite{Loomis1972TheSuperalloys} & 927 & 100 & - & - & - & - \\
51 & \cite{Loomis1972TheSuperalloys} & 927 & 100 & - & - & - & - \\
52 & \cite{Loomis1972TheSuperalloys} & 927 & 100 & - & - & - & - \\
53 & \cite{Loomis1972TheSuperalloys} & 927 & 100 & - & - & - & - \\
54 & \cite{Loomis1972TheSuperalloys} & 927 & 100 & - & - & - & - \\
55 & \cite{Loomis1972TheSuperalloys} & 927 & 100 & - & - & - & - \\
56 & \cite{Loomis1972TheSuperalloys} & 927 & 100 & - & - & - & - \\
57 & \cite{Loomis1972TheSuperalloys} & 927 & 100 & - & - & - & - \\
58 & \cite{Loomis1972TheSuperalloys} & 927 & 100 & - & - & - & - \\
59 & \cite{Blavette1986AnSuperalloys} & 1100 & 4 & 850 & 48 & - & - \\
60 & \cite{Blavette1986AnSuperalloys} & 1100 & 4 & 850 & 48 & - & - \\
61 & \cite{Blavette1986AnSuperalloys} & 1100 & 4 & 850 & 48 & - & - \\
62 & \cite{Blavette1986AnSuperalloys} & 1100 & 4 & 850 & 48 & - & - \\
63 & \cite{Reed2004IdentificationTomography,Reed2004IdentificationTomographyb} & 1140 & 5 & - & - & - & - \\
64 & \cite{Reed2004IdentificationTomography,Reed2004IdentificationTomographyb} & 1140 & 5 & - & - & - & - \\
65 & \cite{Delargy1983PhaseIN939} & 1000 & 6 & 900 & 24 & 700 & 16 \\
66 & \cite{Ralph1982TheTechniques} & - & - & - & - & - & - \\
67 & \cite{Blavette1996Atomic-scaleSuperalloys} & - & - & - & - & - & - \\
68 & \cite{Yoon2007EffectsObservations} & 980 & 0.5 & - & - & - & - \\
69 & \cite{Yoon2007EffectsObservations} & 980 & 0.5 & 800 & 0.25 & - & - \\
70 & \cite{Yoon2007EffectsObservations} & 980 & 0.5 & 800 & 1 & - & - \\
71 & \cite{Yoon2007EffectsObservations} & 980 & 0.5 & 800 & 4 & - & - \\
72 & \cite{Yoon2007EffectsObservations} & 980 & 0.5 & 800 & 16 & - & - \\
73 & \cite{Yoon2007EffectsObservations} & 980 & 0.5 & 800 & 64 & - & - \\
74 & \cite{Yoon2007EffectsObservations} & 980 & 0.5 & 800 & 264 & - & - \\
75 & \cite{Parsa2015AdvancedSuperalloys} & 1140 & 4 & 870 & 16 & - & - \\
76 & \cite{Ma2007DevelopmentSuperalloys} & 1100 & 8 & - & - & - & - \\
77 & \cite{Ma2007DevelopmentSuperalloys} & 1100 & 8 & - & - & - & - \\
78 & \cite{Ma2007DevelopmentSuperalloys} & 1100 & 8 & - & - & - & - \\
79 & \cite{Ma2007DevelopmentSuperalloys} & 1100 & 8 & - & - & - & - \\
80 & \cite{Collier1986EffectsProperties.} & 1066 & 4 & 760 & 16 & - & - \\
81 & \cite{Collier1986EffectsProperties.} & 1066 & 4 & 760 & 16 & - & - \\
82 & \cite{Collier1986EffectsProperties.} & - & - & - & - & - & - \\
83 & \cite{Collier1986EffectsProperties.} & - & - & - & - & - & - \\
84 & \cite{Collier1986EffectsProperties.} & 843 & 24 & - & - & - & - \\
85 & \cite{Collier1986EffectsProperties.} & 843 & 24 & - & - & - & - \\
86 & \cite{Collier1986EffectsProperties.} & 843 & 24 & - & - & - & - \\
87 & \cite{Collier1986EffectsProperties.} & 843 & 24 & - & - & - & - \\
88 & \cite{Long2018MicrostructuralReview} & 1130 & 4 & 900 & 16 & - & - \\
89 & \cite{Wlodek1996TheDT} & 760 & 8 & - & - & - & - \\
90 & \cite{Lapington2018CharacterizationSuperalloys} & 850 & 4 & - & - & - & - \\
91 & \cite{Lapington2018CharacterizationSuperalloys} & 850 & 4 & - & - & - & - \\
92 & \cite{Lapington2018CharacterizationSuperalloys} & 850 & 4 & - & - & - & - \\
93 & \cite{Lapington2018CharacterizationSuperalloys} & 850 & 4 & - & - & - & - \\
94* & \cite{Sulzer2020TheApplications} & 1120 & 4 & 845 & 24 & - & - \\
95* & \cite{Sulzer2020TheApplications} & 1120 & 4 & 845 & 24 & - & - \\
96* & \cite{Sulzer2020TheApplications} & 1120 & 4 & 845 & 24 & - & - \\
97* & \cite{Sulzer2020TheApplications} & 1120 & 4 & 845 & 24 & - & -
\\ \hline
\end{longtable}

\begin{longtable}{llrrrrrrrrrrrr}
\hline
\multicolumn{1}{c}{\#} & \multicolumn{1}{c}{Ref.} & \multicolumn{12}{c}{log at. $\gamma/\gamma'$   partitioning coefficient $P$} \\
\multicolumn{1}{c}{} & \multicolumn{1}{c}{} & \multicolumn{1}{c}{Ni} & \multicolumn{1}{c}{Cr} & \multicolumn{1}{c}{Co} & \multicolumn{1}{c}{Re} & \multicolumn{1}{c}{Ru} & \multicolumn{1}{c}{Al} & \multicolumn{1}{c}{Ta} & \multicolumn{1}{c}{W} & \multicolumn{1}{c}{Ti} & \multicolumn{1}{c}{Mo} & \multicolumn{1}{c}{Nb} & \multicolumn{1}{c}{Hf} \\ \hline\hline
1 & \cite{Duval1994PhaseMC2} & -0.250 & 2.666 & 0.956 & - & - & -1.862 & -1.992 & 0.280 & -2.335 & 1.540 & - & - \\
2 & \cite{Duval1994PhaseMC2} & -0.189 & 2.188 & 0.752 & - & - & -1.120 & -1.792 & 0.460 & -1.735 & 0.938 & - & - \\
3 & \cite{Duval1994PhaseMC2} & -0.154 & 1.689 & 0.610 & - & - & -0.544 & -0.894 & 0.405 & -0.693 & 1.041 & - & - \\
4 & \cite{Duval1994PhaseMC2} & -0.185 & 2.064 & 0.727 & - & - & -0.882 & -1.344 & 0.365 & -1.417 & 1.273 & - & - \\
5 & \cite{Glas1996OrderSuperalloys} & -0.326 & 2.946 & 1.454 & - & - & -1.497 & -1.253 & -0.055 & -2.199 & 1.398 & - & - \\
6 & \cite{Harada1993Atom-probeSuperalloy} & 0.016 & 1.577 & - & - & - & -0.962 & -5.617 & - & - & 0.795 & - & - \\
7 & \cite{Harada1988PhaseSuperalloys} & -0.146 & 1.785 & 0.905 & - & - & -1.424 & -1.386 & 0.392 & - & - & - & - \\
8 & \cite{Harada1988PhaseSuperalloys} & -0.118 & 2.030 & 1.039 & - & - & -1.627 & -1.331 & 0.916 & -1.350 & - & - & - \\
9 & \cite{Harada1988PhaseSuperalloys} & -0.214 & 2.174 & 1.177 & - & - & -1.529 & -1.081 & -0.115 & - & 1.262 & - & - \\
10 & \cite{Harada1988PhaseSuperalloys,Jalilvand2013InfluenceIN-738LC} & -0.267 & 2.109 & 1.125 & - & - & -1.394 & -1.504 & 0.588 & -1.665 & 1.609 & -1.099 & - \\
11 & \cite{Harada1988PhaseSuperalloys,Collier1986EffectsProperties.} & -0.101 & 1.471 & 1.008 & - & - & -1.276 & -1.504 & - & -1.386 & 0.759 & - & -1.792 \\
12 & \cite{Harada1988PhaseSuperalloys,Basak2017MicrostructureSLE} & -0.167 & 1.668 & 0.929 & - & - & -1.146 & -1.386 & 0.134 & -1.447 & 0.916 & - & -1.792 \\
13 & \cite{Harada1988PhaseSuperalloys} & -0.130 & 2.285 & 0.878 & - & - & -1.672 & -5.704 & 0.168 & -2.683 & 1.311 & - & - \\
14 & \cite{Khan1986EffectCMSX-2} & -0.132 & 2.307 & 0.879 & - & - & -1.697 & -5.711 & 0.214 & -3.307 & 1.416 & - & - \\
15 & \cite{Khan1986EffectCMSX-2} & -0.132 & 2.307 & 0.879 & - & - & -1.697 & -5.711 & 0.214 & -3.307 & 1.416 & - & - \\
16 & \cite{Khan1984TheSuperalloy} & -0.086 & 2.187 & - & - & - & -1.849 & -0.955 & 0.162 & - & 1.594 & - & - \\
17 & \cite{Khan1984TheSuperalloy} & -0.218 & 2.211 & 1.163 & - & - & -1.561 & -1.210 & -0.145 & - & 1.456 & - & - \\
18 & \cite{Khan1984TheSuperalloy} & -0.266 & 2.371 & 1.129 & - & - & -2.598 & -1.135 & -0.039 & - & 1.350 & - & - \\
19 & \cite{Miller1994APFIMSuperalloy} & -0.270 & 2.444 & 1.009 & - & - & -1.478 & -1.232 & 0.916 & -1.569 & - & - & - \\
20 & \cite{Miller1994APFIMSuperalloy} & -0.226 & 2.929 & 1.369 & - & - & -1.735 & -1.792 & 1.329 & -2.037 & - & - & - \\
21 & \cite{Royer1998InSuperalloy} & -0.289 & 2.520 & 1.665 & - & - & -1.623 & -1.665 & 0.140 & -1.658 & 1.273 & - & - \\
22 & \cite{Diologent2004OnSuperalloys} & -0.244 & 2.711 & 0.850 & - & - & -2.063 & -1.598 & 0.311 & -1.509 & 1.239 & - & - \\
23 & \cite{Diologent2004OnSuperalloys} & -0.244 & 2.711 & 0.850 & - & - & -2.063 & -1.598 & 0.311 & -1.509 & 1.239 & - & - \\
24 & \cite{Diologent2004OnSuperalloys} & -0.244 & 2.711 & 0.850 & - & - & -2.063 & -1.598 & 0.311 & -1.509 & 1.239 & - & - \\
25 & \cite{Segersall2015Thermal-mechanicalAlloying} & -0.278 & 2.775 & 1.358 & - & - & -1.656 & -3.201 & 0.673 & - & 1.738 & - & -2.485 \\
26 & \cite{Segersall2015Thermal-mechanicalAlloying} & -0.083 & 2.551 & 1.555 & 3.834 & - & -1.860 & -3.511 & 0.217 & - & 1.379 & - & -0.288 \\
27 & \cite{Schmidt1992Effect99} & -0.167 & 2.025 & 1.077 & - & - & -2.419 & -4.868 & 0.520 & -2.688 & - & - & - \\
28 & \cite{Schmidt1992Effect99} & -0.186 & 2.315 & 1.238 & - & - & -3.458 & -4.868 & 0.610 & -3.977 & - & - & - \\
29 & \cite{Schmidt1992Effect99} & -0.220 & 2.314 & 1.343 & - & - & -6.733 & -4.868 & 0.669 & -5.220 & - & - & - \\
30 & \cite{Schmidt1992Effect99} & -0.201 & 1.969 & 1.085 & - & - & -1.048 & -1.329 & 0.373 & -1.329 & - & - & - \\
31 & \cite{Loomis1972TheSuperalloys} & 0.027 & 0.763 & - & - & - & -0.985 & - & - & - & - & - & - \\
32 & \cite{Loomis1972TheSuperalloys} & -0.001 & 1.322 & - & - & - & -1.123 & - & - & - & -0.463 & - & - \\
33 & \cite{Loomis1972TheSuperalloys} & 0.000 & 1.331 & - & - & - & -1.239 & - & - & - & -0.033 & - & - \\
34 & \cite{Loomis1972TheSuperalloys} & -0.038 & 1.508 & - & - & - & -1.307 & - & - & - & 0.397 & - & - \\
35 & \cite{Loomis1972TheSuperalloys} & 0.006 & 0.787 & - & - & - & -0.846 & - & - & - & - & - & - \\
36 & \cite{Loomis1972TheSuperalloys} & -0.023 & 1.124 & - & - & - & -0.900 & - & - & - & -0.324 & - & - \\
37 & \cite{Loomis1972TheSuperalloys} & -0.042 & 1.448 & - & - & - & -1.126 & - & - & - & 0.083 & - & - \\
38 & \cite{Loomis1972TheSuperalloys} & -0.088 & 1.768 & - & - & - & -1.130 & - & - & - & 0.367 & - & - \\
39 & \cite{Loomis1972TheSuperalloys} & -0.050 & 0.927 & - & - & - & -0.561 & - & - & - & - & - & - \\
40 & \cite{Loomis1972TheSuperalloys} & -0.111 & 1.455 & - & - & - & -0.610 & - & - & - & -0.222 & - & - \\
41 & \cite{Loomis1972TheSuperalloys} & -0.151 & 2.074 & - & - & - & -0.832 & - & - & - & 0.161 & - & - \\
42 & \cite{Loomis1972TheSuperalloys} & -0.145 & 1.545 & - & - & - & -0.704 & - & - & - & 0.333 & - & - \\
43 & \cite{Loomis1972TheSuperalloys} & 0.037 & 2.039 & - & - & - & -1.864 & - & - & -1.595 & - & - & - \\
44 & \cite{Loomis1972TheSuperalloys} & 0.028 & 2.132 & - & - & - & -1.799 & - & - & -1.736 & 1.211 & - & - \\
45 & \cite{Loomis1972TheSuperalloys} & -0.031 & 2.567 & - & - & - & -1.780 & - & - & -1.553 & 1.677 & - & - \\
46 & \cite{Loomis1972TheSuperalloys} & -0.044 & 2.635 & - & - & - & -1.977 & - & - & -1.637 & 1.804 & - & - \\
47 & \cite{Loomis1972TheSuperalloys} & 0.055 & 0.595 & - & - & - & -0.840 & - & - & - & - & - & - \\
48 & \cite{Loomis1972TheSuperalloys} & 0.015 & 0.779 & - & - & - & -0.791 & - & - & - & 0.179 & - & - \\
49 & \cite{Loomis1972TheSuperalloys} & -0.041 & 1.092 & - & - & - & -0.721 & - & - & - & 0.338 & - & - \\
50 & \cite{Loomis1972TheSuperalloys} & -0.071 & 1.342 & - & - & - & -0.764 & - & - & - & 0.507 & - & - \\
51 & \cite{Loomis1972TheSuperalloys} & -0.019 & 0.708 & - & - & - & -0.488 & - & - & - & - & - & - \\
52 & \cite{Loomis1972TheSuperalloys} & -0.050 & 1.015 & - & - & - & -0.549 & - & - & - & -0.060 & - & - \\
53 & \cite{Loomis1972TheSuperalloys} & -0.102 & 1.341 & - & - & - & -0.494 & - & - & - & 0.185 & - & - \\
54 & \cite{Loomis1972TheSuperalloys} & -0.107 & 1.390 & - & - & - & -0.564 & - & - & - & 0.417 & - & - \\
55 & \cite{Loomis1972TheSuperalloys} & 0.019 & 2.002 & - & - & - & -1.411 & - & - & -1.227 & - & - & - \\
56 & \cite{Loomis1972TheSuperalloys} & 0.017 & 1.950 & - & - & - & -1.474 & - & - & -1.281 & 1.478 & - & - \\
57 & \cite{Loomis1972TheSuperalloys} & 0.023 & 2.256 & - & - & - & -1.500 & - & - & -1.443 & 1.689 & - & - \\
58 & \cite{Loomis1972TheSuperalloys} & -0.019 & 2.629 & - & - & - & -1.675 & - & - & -1.369 & 1.874 & - & - \\
59 & \cite{Blavette1986AnSuperalloys} & -0.353 & 2.753 & 1.267 & - & - & -1.708 & -2.104 & 0.247 & -2.819 & - & - & - \\
60 & \cite{Blavette1986AnSuperalloys} & -0.381 & 2.696 & 1.340 & 1.585 & - & -1.799 & -2.839 & - & -2.299 & - & - & - \\
61 & \cite{Blavette1986AnSuperalloys} & -0.168 & 2.363 & 0.989 & - & - & -1.684 & -3.401 & 0.041 & -0.981 & - & - & - \\
62 & \cite{Blavette1986AnSuperalloys} & -0.148 & 1.916 & 0.843 & 1.946 & - & -1.314 & -1.749 & 0.704 & -2.096 & - & - & - \\
63 & \cite{Reed2004IdentificationTomography,Reed2004IdentificationTomographyb} & -0.311 & 1.930 & 1.138 & 2.836 & - & -1.699 & -2.347 & 0.236 & - & - & - & - \\
64 & \cite{Reed2004IdentificationTomography,Reed2004IdentificationTomographyb} & -0.362 & 1.959 & 1.109 & 2.821 & 1.193 & -1.681 & -2.383 & 0.196 & - & - & - & - \\
65 & \cite{Delargy1983PhaseIN939} & -0.456 & 2.999 & 0.706 & - & - & -2.343 & - & - & -2.906 & - & - & - \\
66 & \cite{Ralph1982TheTechniques} & 0.133 & - & - & - & - & -0.609 & - & - & -0.585 & - & - & - \\
67 & \cite{Blavette1996Atomic-scaleSuperalloys} & -0.294 & 2.385 & 0.611 & - & - & -1.337 & - & - & -2.503 & 0.797 & - & - \\
68 & \cite{Yoon2007EffectsObservations} & 0.056 & 0.404 & - & 0.142 & - & -0.700 & - & - & - & - & - & - \\
69 & \cite{Yoon2007EffectsObservations} & 0.057 & 0.578 & - & 0.487 & - & -0.827 & - & - & - & - & - & - \\
70 & \cite{Yoon2007EffectsObservations} & 0.052 & 0.681 & - & 0.758 & - & -0.887 & - & - & - & - & - & - \\
71 & \cite{Yoon2007EffectsObservations} & 0.054 & 0.677 & - & 0.777 & - & -0.914 & - & - & - & - & - & - \\
72 & \cite{Yoon2007EffectsObservations} & 0.054 & 0.691 & - & 0.841 & - & -0.916 & - & - & - & - & - & - \\
73 & \cite{Yoon2007EffectsObservations} & 0.060 & 0.669 & - & 0.847 & - & -0.944 & - & - & - & - & - & - \\
74 & \cite{Yoon2007EffectsObservations} & 0.060 & 0.649 & - & 0.925 & - & -0.943 & - & - & - & - & - & - \\
75 & \cite{Parsa2015AdvancedSuperalloys} & -0.393 & 2.371 & 1.207 & 3.134 & - & -1.819 & -3.095 & 0.638 & -2.525 & 1.212 & - & -1.209 \\
76 & \cite{Ma2007DevelopmentSuperalloys} & -0.257 & 1.701 & 0.722 & 6.367 & 1.218 & -1.058 & -2.312 & 0.595 & - & - & - & - \\
77 & \cite{Ma2007DevelopmentSuperalloys} & -0.375 & 1.783 & 0.807 & 4.474 & 1.169 & -1.195 & -2.011 & 0.636 & - & - & - & - \\
78 & \cite{Ma2007DevelopmentSuperalloys} & -0.400 & 2.036 & 0.839 & 7.029 & 1.508 & -1.341 & -1.765 & 0.769 & - & - & - & - \\
79 & \cite{Ma2007DevelopmentSuperalloys} & -0.544 & 1.888 & 0.780 & 7.078 & 1.245 & -1.299 & -2.105 & 0.931 & - & - & - & - \\
80 & \cite{Collier1986EffectsProperties.} & -0.297 & 1.845 & 0.714 & - & - & -2.040 & - & -0.111 & -1.895 & 1.269 & 0.022 & - \\
81 & \cite{Collier1986EffectsProperties.} & -0.304 & 2.072 & 0.768 & - & - & -3.346 & - & - & -2.059 & 2.006 & 0.010 & - \\
82 & \cite{Collier1986EffectsProperties.} & -0.159 & 1.659 & 1.016 & - & - & -6.984 & -1.435 & - & -2.792 & 1.715 & 1.173 & - \\
83 & \cite{Collier1986EffectsProperties.} & -0.039 & 1.128 & 0.733 & - & - & -2.614 & - & - & -0.340 & 0.941 & 1.454 & - \\
84 & \cite{Collier1986EffectsProperties.} & -0.264 & 1.812 & 1.117 & - & - & -1.936 & 0.489 & 0.542 & -1.582 & 1.385 & 0.168 & - \\
85 & \cite{Collier1986EffectsProperties.} & -0.185 & 2.098 & 1.279 & - & - & -6.768 & -0.161 & - & -2.005 & 1.068 & 1.103 & - \\
86 & \cite{Collier1986EffectsProperties.} & -0.184 & 2.183 & 1.303 & - & - & -6.847 & - & 0.486 & -1.659 & 1.095 & 0.981 & - \\
87 & \cite{Collier1986EffectsProperties.} & -0.168 & 1.778 & 1.169 & - & - & -3.063 & -0.130 & 0.683 & -1.745 & 0.873 & 0.747 & - \\
88 & \cite{Long2018MicrostructuralReview} & -0.113 & 2.072 & -0.208 & 1.641 & - & -1.045 & -1.407 & -0.307 & - & 0.787 & - & - \\
89 & \cite{Wlodek1996TheDT} & -0.347 & 2.398 & 1.273 & - & - & -1.694 & - & 0.213 & -2.505 & 1.658 & -1.625 & - \\
90 & \cite{Lapington2018CharacterizationSuperalloys} & -0.692 & 2.928 & 1.151 & - & - & -1.694 & -2.485 & 0.000 & -3.308 & - & - & - \\
91 & \cite{Lapington2018CharacterizationSuperalloys} & -0.647 & 2.827 & 1.133 & - & - & -1.580 & -2.303 & 0.201 & -2.862 & - & -1.792 & - \\
92 & \cite{Lapington2018CharacterizationSuperalloys} & -0.576 & 2.733 & 1.122 & - & - & -1.708 & -2.398 & 0.000 & -2.741 & - & -1.872 & - \\
93 & \cite{Lapington2018CharacterizationSuperalloys} & -0.652 & 2.816 & 1.079 & - & - & -1.705 & -2.485 & 0.452 & -2.944 & - & -1.846 & - \\
94* & \cite{Sulzer2020TheApplications} & -0.333 & 2.492 & 1.294 & - & - & -1.678 & -3.559 & -0.411 & - & 1.067 & - & - \\
95* & \cite{Sulzer2020TheApplications} & -0.313 & 2.729 & 1.449 & - & - & -1.933 & -3.515 & 0.174 & - & 1.327 & - & -3.466 \\
96* & \cite{Sulzer2020TheApplications} & -0.431 & 3.086 & 1.435 & - & - & -1.699 & -3.784 & 0.561 & -3.013 & 1.680 & - & - \\
97* & \cite{Sulzer2020TheApplications} & -0.250 & 2.531 & 0.520 & - & - & -1.953 & -3.250 & -0.070 & - & 1.583 & - & - \\ \hline
\end{longtable}

\begin{longtable}{llrrrrrrrrrrrrr}
\hline
\# & Ref. & \multicolumn{12}{c}{log at. nom./$\gamma'$   partitioning coefficient $P'$} & \multicolumn{1}{c}{$\gamma'$ at. fraction} \\
 &  & \multicolumn{1}{c}{Ni} & \multicolumn{1}{c}{Cr} & \multicolumn{1}{c}{Co} & \multicolumn{1}{c}{Re} & \multicolumn{1}{c}{Ru} & \multicolumn{1}{c}{Al} & \multicolumn{1}{c}{Ta} & \multicolumn{1}{c}{W} & \multicolumn{1}{c}{Ti} & \multicolumn{1}{c}{Mo} & \multicolumn{1}{c}{Nb} & \multicolumn{1}{c}{Hf} & \multicolumn{1}{c}{} \\ \hline\hline
1 & \cite{Duval1994PhaseMC2} & -0.073 & 1.642 & 0.376 & - & - & -0.279 & -0.095 & -0.176 & -0.215 & 0.773 & - & - & 0.690 \\
2 & \cite{Duval1994PhaseMC2} & -0.062 & 1.355 & 0.218 & - & - & -0.251 & -0.182 & 0.080 & -0.307 & 0.368 & - & - & 0.640 \\
3 & \cite{Duval1994PhaseMC2} & -0.063 & 1.165 & 0.148 & - & - & -0.244 & -0.095 & 0.080 & -0.182 & 0.773 & - & - & 0.480 \\
4 & \cite{Duval1994PhaseMC2} & -0.043 & 1.314 & 0.148 & - & - & -0.338 & -0.140 & 0.039 & -0.278 & 0.619 & - & - & 0.510 \\
5 & \cite{Glas1996OrderSuperalloys} & -0.084 & 1.829 & 0.668 & - & - & -0.255 & -0.232 & -0.016 & -0.298 & 0.633 & - & - & 0.710 \\
6 & \cite{Harada1993Atom-probeSuperalloy} & 0.008 & 0.956 & - & - & - & -0.280 & -0.675 & - & - & 0.427 & - & - & 0.580 \\
7 & \cite{Harada1988PhaseSuperalloys} & -0.045 & 0.953 & 0.389 & - & - & -0.276 & -0.289 & 0.141 & - & - & - & - & 0.640 \\
8 & \cite{Harada1988PhaseSuperalloys} & -0.056 & 1.326 & 0.575 & - & - & -0.401 & -0.275 & 0.502 & -0.350 & - & - & - & 0.590 \\
9 & \cite{Harada1988PhaseSuperalloys} & -0.062 & 1.229 & 0.528 & - & - & -0.278 & -0.229 & -0.034 & - & 0.580 & - & - & 0.690 \\
10 & \cite{Harada1988PhaseSuperalloys,Jalilvand2013InfluenceIN-738LC} & -0.156 & 1.678 & 0.773 & - & - & -0.584 & -0.518 & 0.479 & -0.602 & 1.272 & -0.487 & - & 0.420 \\
11 & \cite{Harada1988PhaseSuperalloys,Collier1986EffectsProperties.} & -0.045 & 0.994 & 0.588 & - & - & -0.469 & -0.368 & - & -0.039 & 0.468 & - & - & 0.520 \\
12 & \cite{Harada1988PhaseSuperalloys,Basak2017MicrostructureSLE} & -0.083 & 1.022 & 0.457 & - & - & -0.267 & -0.194 & 0.146 & -0.312 & 0.776 & - & -0.586 & 0.590 \\
13 & \cite{Harada1988PhaseSuperalloys} & -0.039 & 1.343 & 0.376 & - & - & -0.298 & -0.449 & 0.067 & -0.354 & 0.625 & - & - & 0.680 \\
14 & \cite{Khan1986EffectCMSX-2} & -0.040 & 1.361 & 0.378 & - & - & -0.300 & -0.457 & 0.085 & -0.369 & 0.692 & - & - & 0.680 \\
15 & \cite{Khan1986EffectCMSX-2} & -0.040 & 1.361 & 0.378 & - & - & -0.300 & -0.457 & 0.085 & -0.369 & 0.692 & - & - & 0.680 \\
16 & \cite{Khan1984TheSuperalloy} & -0.027 & 1.263 & - & - & - & -0.315 & -0.220 & 0.055 & - & 0.815 & - & - & 0.679 \\
17 & \cite{Khan1984TheSuperalloy} & -0.062 & 1.253 & 0.517 & - & - & -0.279 & -0.243 & -0.043 & - & 0.699 & - & - & 0.692 \\
18 & \cite{Khan1984TheSuperalloy} & -0.071 & 1.351 & 0.481 & - & - & -0.319 & -0.223 & -0.011 & - & 0.611 & - & - & 0.705 \\
19 & \cite{Miller1994APFIMSuperalloy} & -0.116 & 1.609 & 0.376 & - & - & -0.303 & 0.511 & 0.000 & -0.234 & - & - & - & 0.700 \\
20 & \cite{Miller1994APFIMSuperalloy} & -0.048 & 2.106 & 0.565 & - & - & -0.435 & -0.588 & 0.442 & -0.191 & - & - & - & 0.700 \\
21 & \cite{Royer1998InSuperalloy} & -0.094 & 1.421 & 1.145 & - & - & -0.253 & -0.315 & -0.105 & -0.336 & 0.619 & - & - & 0.700 \\
22 & \cite{Diologent2004OnSuperalloys} & -0.067 & 1.651 & 0.338 & - & - & -0.304 & -0.274 & 0.104 & -0.266 & 0.552 & - & - & 0.700 \\
23 & \cite{Diologent2004OnSuperalloys} & -0.067 & 1.651 & 0.338 & - & - & -0.304 & -0.274 & 0.104 & -0.266 & 0.552 & - & - & 0.700 \\
24 & \cite{Diologent2004OnSuperalloys} & -0.067 & 1.651 & 0.338 & - & - & -0.304 & -0.274 & 0.104 & -0.266 & 0.552 & - & - & 0.700 \\
25 & \cite{Segersall2015Thermal-mechanicalAlloying} & -0.143 & 2.208 & 0.948 & - & - & -0.526 & -0.792 & 0.508 & - & 1.355 & - & -1.386 & 0.460 \\
26 & \cite{Segersall2015Thermal-mechanicalAlloying} & -0.010 & 2.227 & 1.067 & 2.773 & - & -0.916 & -0.879 & 0.047 & - & 0.527 & - & 0.000 & 0.320 \\
27 & \cite{Schmidt1992Effect99} & -0.045 & 1.068 & 0.446 & - & - & -0.306 & -0.368 & 0.182 & -0.315 & - & - & - & 0.710 \\
28 & \cite{Schmidt1992Effect99} & -0.051 & 1.306 & 0.545 & - & - & -0.336 & -0.368 & 0.223 & -0.342 & - & - & - & 0.705 \\
29 & \cite{Schmidt1992Effect99} & -0.051 & 1.232 & 0.580 & - & - & -0.336 & -0.368 & 0.266 & -0.315 & - & - & - & 0.730 \\
30 & \cite{Schmidt1992Effect99} & -0.064 & 1.130 & 0.511 & - & - & -0.249 & -0.288 & 0.143 & -0.288 & - & - & - & 0.660 \\
31 & \cite{Loomis1972TheSuperalloys} & 0.026 & 0.747 & - & - & - & -0.936 & - & - & - & - & - & - & 0.030 \\
32 & \cite{Loomis1972TheSuperalloys} & -0.001 & 1.246 & - & - & - & -0.935 & - & - & - & -0.405 & - & - & 0.100 \\
33 & \cite{Loomis1972TheSuperalloys} & 0.000 & 1.222 & - & - & - & -0.944 & - & - & - & -0.029 & - & - & 0.140 \\
34 & \cite{Loomis1972TheSuperalloys} & -0.031 & 1.348 & - & - & - & -0.894 & - & - & - & 0.333 & - & - & 0.190 \\
35 & \cite{Loomis1972TheSuperalloys} & 0.005 & 0.684 & - & - & - & -0.632 & - & - & - & - & - & - & 0.180 \\
36 & \cite{Loomis1972TheSuperalloys} & -0.018 & 0.955 & - & - & - & -0.611 & - & - & - & -0.240 & - & - & 0.230 \\
37 & \cite{Loomis1972TheSuperalloys} & -0.029 & 1.188 & - & - & - & -0.640 & - & - & - & 0.059 & - & - & 0.300 \\
38 & \cite{Loomis1972TheSuperalloys} & -0.058 & 1.449 & - & - & - & -0.604 & - & - & - & 0.260 & - & - & 0.330 \\
39 & \cite{Loomis1972TheSuperalloys} & -0.033 & 0.697 & - & - & - & -0.333 & - & - & - & - & - & - & 0.340 \\
40 & \cite{Loomis1972TheSuperalloys} & -0.063 & 1.066 & - & - & - & -0.308 & - & - & - & -0.123 & - & - & 0.420 \\
41 & \cite{Loomis1972TheSuperalloys} & -0.077 & 1.545 & - & - & - & -0.356 & - & - & - & 0.088 & - & - & 0.470 \\
42 & \cite{Loomis1972TheSuperalloys} & -0.073 & 1.071 & - & - & - & -0.305 & - & - & - & 0.187 & - & - & 0.480 \\
43 & \cite{Loomis1972TheSuperalloys} & 0.033 & 1.929 & - & - & - & -1.361 & - & - & -1.209 & - & - & - & 0.120 \\
44 & \cite{Loomis1972TheSuperalloys} & 0.025 & 2.010 & - & - & - & -1.295 & - & - & -1.261 & 1.116 & - & - & 0.130 \\
45 & \cite{Loomis1972TheSuperalloys} & -0.026 & 2.418 & - & - & - & -1.226 & - & - & -1.109 & 1.547 & - & - & 0.150 \\
46 & \cite{Loomis1972TheSuperalloys} & -0.037 & 2.474 & - & - & - & -1.286 & - & - & -1.129 & 1.660 & - & - & 0.160 \\
47 & \cite{Loomis1972TheSuperalloys} & 0.052 & 0.572 & - & - & - & -0.776 & - & - & - & - & - & - & 0.050 \\
48 & \cite{Loomis1972TheSuperalloys} & 0.014 & 0.735 & - & - & - & -0.699 & - & - & - & 0.166 & - & - & 0.080 \\
49 & \cite{Loomis1972TheSuperalloys} & -0.035 & 1.002 & - & - & - & -0.592 & - & - & - & 0.300 & - & - & 0.130 \\
50 & \cite{Loomis1972TheSuperalloys} & -0.058 & 1.208 & - & - & - & -0.586 & - & - & - & 0.437 & - & - & 0.170 \\
51 & \cite{Loomis1972TheSuperalloys} & -0.015 & 0.601 & - & - & - & -0.370 & - & - & - & - & - & - & 0.200 \\
52 & \cite{Loomis1972TheSuperalloys} & -0.036 & 0.810 & - & - & - & -0.357 & - & - & - & -0.043 & - & - & 0.290 \\
53 & \cite{Loomis1972TheSuperalloys} & -0.067 & 1.062 & - & - & - & -0.302 & - & - & - & 0.128 & - & - & 0.330 \\
54 & \cite{Loomis1972TheSuperalloys} & -0.071 & 1.115 & - & - & - & -0.347 & - & - & - & 0.301 & - & - & 0.320 \\
55 & \cite{Loomis1972TheSuperalloys} & 0.018 & 1.975 & - & - & - & -1.322 & - & - & -1.157 & - & - & - & 0.030 \\
56 & \cite{Loomis1972TheSuperalloys} & 0.017 & 1.915 & - & - & - & -1.348 & - & - & -1.182 & 1.447 & - & - & 0.040 \\
57 & \cite{Loomis1972TheSuperalloys} & 0.022 & 2.210 & - & - & - & -1.340 & - & - & -1.293 & 1.647 & - & - & 0.050 \\
58 & \cite{Loomis1972TheSuperalloys} & -0.018 & 2.561 & - & - & - & -1.410 & - & - & -1.182 & 1.813 & - & - & 0.070 \\
59 & \cite{Blavette1986AnSuperalloys} & -0.081 & 1.683 & 0.508 & - & - & -0.313 & -0.130 & -0.241 & -0.405 & - & - & - & 0.711 \\
60 & \cite{Blavette1986AnSuperalloys} & -0.079 & 1.621 & 0.581 & 0.544 & - & -0.310 & -0.336 & - & -0.486 & - & - & - & 0.719 \\
61 & \cite{Blavette1986AnSuperalloys} & -0.046 & 1.346 & 0.462 & - & - & -0.313 & -0.410 & 0.084 & -0.247 & - & - & - & 0.680 \\
62 & \cite{Blavette1986AnSuperalloys} & -0.039 & 1.049 & 0.441 & 1.146 & - & -0.344 & -0.407 & 0.368 & -0.403 & - & - & - & 0.662 \\
63 & \cite{Reed2004IdentificationTomography,Reed2004IdentificationTomographyb} & -0.063 & 0.777 & 0.401 & 1.609 & - & -0.188 & -0.278 & 0.081 & - & - & - & - & 0.773 \\
64 & \cite{Reed2004IdentificationTomography,Reed2004IdentificationTomographyb} & -0.076 & 0.806 & 0.391 & 1.543 & 0.390 & -0.153 & -0.355 & 0.046 & - & - & - & - & 0.776 \\
65 & \cite{Delargy1983PhaseIN939} & -0.261 & 2.610 & 0.469 & - & - & -1.144 & - & - & -1.075 & - & - & - & 0.344 \\
66 & \cite{Ralph1982TheTechniques} & 0.128 & - & - & - & - & -0.564 & - & - & -0.585 & - & - & - & 0.040 \\
67 & \cite{Blavette1996Atomic-scaleSuperalloys} & -0.160 & 1.905 & 0.397 & - & - & -0.558 & - & - & -0.760 & 0.535 & - & - & 0.420 \\
68 & \cite{Yoon2007EffectsObservations} & 0.043 & 0.274 & - & 0.122 & - & -0.445 & - & - & - & - & - & - & 0.180 \\
69 & \cite{Yoon2007EffectsObservations} & 0.041 & 0.443 & - & 0.439 & - & -0.526 & - & - & - & - & - & - & 0.203 \\
70 & \cite{Yoon2007EffectsObservations} & 0.035 & 0.523 & - & 0.693 & - & -0.542 & - & - & - & - & - & - & 0.229 \\
71 & \cite{Yoon2007EffectsObservations} & 0.036 & 0.520 & - & 0.717 & - & -0.547 & - & - & - & - & - & - & 0.247 \\
72 & \cite{Yoon2007EffectsObservations} & 0.036 & 0.533 & - & 0.766 & - & -0.556 & - & - & - & - & - & - & 0.246 \\
73 & \cite{Yoon2007EffectsObservations} & 0.040 & 0.517 & - & 0.760 & - & -0.564 & - & - & - & - & - & - & 0.253 \\
74 & \cite{Yoon2007EffectsObservations} & 0.041 & 0.486 & - & 0.857 & - & -0.566 & - & - & - & - & - & - & 0.254 \\
75 & \cite{Parsa2015AdvancedSuperalloys} & -0.068 & 1.255 & 0.457 & 1.559 & - & -0.269 & -0.357 & 0.148 & -0.318 & 0.424 & - & 0.019 & 0.761 \\
76 & \cite{Ma2007DevelopmentSuperalloys} & -0.101 & 0.796 & -0.037 & - & 0.598 & 0.621 & -1.208 & 0.057 & - & - & - & - & 0.680 \\
77 & \cite{Ma2007DevelopmentSuperalloys} & -0.111 & 0.953 & 0.323 & 3.039 & 0.568 & 0.069 & -1.026 & 0.095 & - & - & - & - & 0.680 \\
78 & \cite{Ma2007DevelopmentSuperalloys} & -0.063 & 1.052 & 0.109 & - & 0.722 & -0.053 & -1.031 & 0.235 & - & - & - & - & 0.680 \\
79 & \cite{Ma2007DevelopmentSuperalloys} & -0.113 & 1.006 & 0.319 & - & 0.545 & -0.020 & -1.037 & 0.234 & - & - & - & - & 0.680 \\
80 & \cite{Collier1986EffectsProperties.} & -0.165 & 1.426 & 0.481 & - & - & -0.725 & - & -0.064 & -0.701 & 0.922 & - & - & 0.407 \\
81 & \cite{Collier1986EffectsProperties.} & -0.170 & 1.637 & 0.524 & - & - & -0.856 & - & - & -0.734 & 1.576 & - & - & 0.404 \\
82 & \cite{Collier1986EffectsProperties.} & -0.028 & 0.844 & 0.465 & - & - & -0.561 & 0.666 & - & -0.440 & 1.033 & - & - & 0.621 \\
83 & \cite{Collier1986EffectsProperties.} & -0.016 & 0.632 & 0.376 & - & - & -0.496 & - & - & -0.130 & 0.506 & - & - & 0.578 \\
84 & \cite{Collier1986EffectsProperties.} & -0.137 & 1.343 & 0.758 & - & - & -0.639 & 0.299 & 0.334 & -0.577 & 0.976 & 0.096 & - & 0.448 \\
85 & \cite{Collier1986EffectsProperties.} & -0.101 & 1.634 & 0.919 & - & - & -0.895 & -0.090 & - & -0.672 & 0.774 & 0.774 & - & 0.420 \\
86 & \cite{Collier1986EffectsProperties.} & -0.098 & 1.708 & 0.944 & - & - & -0.940 & - & 0.311 & -0.577 & 0.861 & 0.679 & - & 0.417 \\
87 & \cite{Collier1986EffectsProperties.} & -0.094 & 1.351 & 0.829 & - & - & -0.809 & -0.073 & 0.451 & -0.655 & 0.594 & - & - & 0.418 \\
88 & \cite{Long2018MicrostructuralReview} & -0.154 & 1.072 & 1.131 & 1.339 & - & -0.062 & -0.628 & 1.116 & - & 0.036 & - & - & 0.626 \\
89 & \cite{Wlodek1996TheDT} & -0.242 & 2.062 & 0.987 & - & - & -0.613 & - & 0.388 & -0.857 & 1.236 & -0.795 & - & 0.311 \\
90 & \cite{Lapington2018CharacterizationSuperalloys} & -0.297 & 2.297 & 0.766 & - & - & -0.543 & -0.539 & 0.223 & -0.646 & - & - & - & 0.487 \\
91 & \cite{Lapington2018CharacterizationSuperalloys} & -0.297 & 2.230 & 0.760 & - & - & -0.543 & -0.511 & 0.000 & -0.560 & - & -0.405 & - & 0.469 \\
92 & \cite{Lapington2018CharacterizationSuperalloys} & -0.276 & 2.181 & 0.755 & - & - & -0.597 & -0.606 & 0.000 & -0.631 & - & -0.486 & - & 0.453 \\
93 & \cite{Lapington2018CharacterizationSuperalloys} & -0.296 & 2.197 & 0.759 & - & - & -0.568 & -0.693 & 0.251 & -0.676 & - & -0.460 & - & 0.471 \\
94* & \cite{Sulzer2020TheApplications} & -0.095 & 1.631 & 0.671 & - & - & -0.449 & -0.078 & -0.446 & - & 0.551 & - & - & 0.592 \\
95* & \cite{Sulzer2020TheApplications} & -0.103 & 1.895 & 0.853 & - & - & -0.410 & -0.720 & 0.990 & - & 0.298 & - & 0.031 & 0.562 \\
96* & \cite{Sulzer2020TheApplications} & -0.111 & 2.091 & 0.758 & - & - & -0.401 & -0.467 & 0.366 & -0.475 & 1.110 & - & - & 0.627 \\
97* & \cite{Sulzer2020TheApplications} & -0.140 & 1.757 & 0.687 & - & - & -0.457 & -0.398 & -0.061 & - & 0.884 & - & - & 0.585 \\ \hline
\end{longtable}
\end{center}
\end{landscape}


\end{document}